\documentclass{aa}  

\usepackage{graphicx}
\usepackage{txfonts}
\usepackage{lipsum}
\usepackage{subcaption}         
\usepackage{lscape}             
\usepackage{placeins}           
                                
\usepackage{hyperref}

\begin{document}

   \title{CAPOS: The bulge Cluster APOgee Survey XII. }

   \subtitle{Abundances for 98 PIGS metal-poor Bulge field giants}


   \author{Carolina Salgado\inst{1}\thanks{Corresponding author: c.salgadoescalona@uandresbello.edu}
        \and Sandro Villanova\inst{1}
        \and Doug Geisler\inst{2, 3}
        \and Nicolás Barrera\inst{3}
        }

\institute{
Universidad Andres Bello, Facultad de Ciencias Exactas, Departamento de Ciencias Físicas – Instituto de Astrofísica,
Autopista Concepción-Talcahuano 7100, Talcahuano, Chile
\and
Departamento de Astronomía, Casilla 160-C, Universidad de Concepción, Concepción, Chile
\and
Departamento de Astronomía, Facultad de Ciencias, Universidad de La Serena, Av. Raul Bitran 1305, La Serena, Chile
}

   \date{}

 
\abstract
{The inner Milky Way hosts overlapping stellar populations (bar--bulge, inner thin and thick disks, and halo), complicating population assignments along the line of sight. A joint chemical--dynamical approach is required to isolate a clean field bulge sample. Metal-poor bulge stars are particularly valuable as they likely trace the earliest phases of chemical enrichment in the inner Galaxy.}
{We aim to characterize the $\alpha$-element (Si, Mg) and selected Fe-peak abundances of a dynamically defined sample of bulge field stars, and to contrast these trends with the inner-halo tail in the metal-poor regime.}
{We analyze metal-poor candidates from the Pristine Inner Galaxy Survey observed by the bulge Cluster APOGEE Survey. Using APOGEE/ASPCAP abundances (${\rm S/N} \ge 50$), we integrate full 6D orbits in a barred Milky Way potential. Bulge membership is defined via orbital confinement using an apocenter cut and the high-density locus in the $(E_J, L_z)$ plane, with uncertainties estimated using Monte Carlo simulations. We identify 98 stars as genuine bulge members.}
{The metallicity distribution spans $-2.5 \le [\mathrm{Fe/H}] \le -0.4$, with a median $[\mathrm{Fe/H}] = -1.71$, offset to higher metallicity than the inner halo (median $[\mathrm{Fe/H}] = -1.96$). The sample follows a high-$\alpha$ sequence with slopes $\mathrm{d}[\mathrm{Si/Fe}]/\mathrm{d}[\mathrm{Fe/H}] = -0.020^{+0.012}_{-0.051}$ and $\mathrm{d}[\mathrm{Mg/Fe}]/\mathrm{d}[\mathrm{Fe/H}] = -0.097^{+0.062}_{-0.133}$. Fe-peak tracers show $[\mathrm{Ni/Fe}] \sim 0$, while $[\mathrm{Mn/Fe}]$ declines with $[\mathrm{Fe/H}]$ with a mild upturn at higher metallicity. We detect no significant $[\mathrm{Si/Fe}]$ gradients with $R_{\rm apo}$ or $Z_{\max}$ ($+0.010^{+0.018}_{0.000}$ and $+0.006^{+0.022}_{-0.011}\ \mathrm{dex\ kpc^{-1}}$, respectively). Results are insensitive to the assumed $\Omega_{\rm p}$, yielding indistinguishable memberships.}
{The chemo-orbital evidence favors an in situ origin within the inner Galaxy for the metal-poor bulge field, enriched at early times and later rearranged by secular bar evolution, with a minor contribution from halo stars at the most metal-poor end.}

   \keywords{Stars: abundances --
                Stars: atmospheres --
                Galaxy: bulge -- 
                Galaxy: kinematics and dynamics
               }

   \maketitle
   \nolinenumbers

\section{Introduction}\label{sec:intro}

The central Milky Way is a composite environment where all major stellar populations overlap along heavily reddened sightlines: a bulge--bar and inner thin and thick disks dominate the mass and kinematics, while a rounder, more metal-poor, dynamically hotter halo component contributes at lower density \citep[e.g.][]{zoccali2008metal, ness2016apogee, barbuy2018chemodynamical, barbuy2025abundances}. Photometric and spectroscopic surveys have established that the bulge is predominantly old (ages $\gtrsim10$\,Gyr), with evidence for a smaller intermediate-age, metal-rich component, pointing to a multi-phase formation history \citep[e.g.][]{bensby2013chemical,clarkson2011first,ness2016apogee,barbuy2018chemodynamical,barbuy2025abundances}. Dynamically, the inner regions exhibit nearly cylindrical rotation and a strong bar signature, while a kinematically hot tail connects to the inner halo and old thick disk \citep[e.g.][]{ness2016apogee,duong2019herbs1,duong2019herbs2,lucey2022combs, babusiaux2010insights, babusiaux2016correlations}. RR~Lyrae studies provide an important complementary view of the oldest bulge populations: BRAVA-RR showed that bulge RR~Lyrae stars exhibit hot kinematics and null or negligible rotation, consistent with a population distinct from the dominant bar/pseudobulge, although either a classical-bulge or metal-rich inner-halo/bulge interpretation remains possible \citep{kunder2016}. Wide-field optical photometry from the Blanco DECam Bulge Survey has mapped more than 200 deg$^2$ of the southern bulge and used near-UV/optical colors to trace red-clump metallicities over large areas \citep{rich2020,johnson2020}. The BDBS metallicity analysis of 2.6 million red clump stars revealed complex MDF morphology, a strong vertical metallicity gradient, and metallicity-dependent structure, with stars at lower metallicity showing a weaker connection to the bar and a more spheroidal or thick-bar-like distribution than the most metal-rich populations \citep{johnson2022}.

Chemical abundances provide a time-axis to this structural view. The $\alpha$ elements (O, Mg, Si, Ca, Ti) are produced promptly in core-collapse SNe\,II, whereas Fe receives an additional delayed  contribution from SNe\,Ia; the shape of the $[\alpha/{\rm Fe}]$--$[\mathrm{Fe/H}]$ sequence therefore traces star-formation rates and time-scales \citep[e.g.][]{tinsley1979stellar,matteucci1990metallicity,mcwilliam2016chemical}. Bulge stars at subsolar metallicity are $\alpha$-enhanced, and the ``knee'' in $[\alpha/{\rm Fe}]$ tends to appear at higher $[\mathrm{Fe/H}]$ than in the solar neighborhood, consistent with rapid early enrichment in the inner Galaxy's deep potential well \citep[e.g.][]{bensby2013chemical,mcwilliam2016chemical,barbuy2018chemodynamical,rojas2020many,barbuy2025abundances}. Differential analyses have further shown that the detailed $\alpha$ patterns of the bulge and the high-$\alpha$ thick disk are similar over $-1.5 \lesssim [\mathrm{Fe/H}] \lesssim -0.3$, favoring scenarios in which much of the present-day bulge originates from the inner disk and was subsequently rearranged by bar-driven evolution rather than formed as a purely ``classical'' bulge \citep[e.g.][]{melendez2008chemical,alves2010chemical,rojas2017gaia,ness2016apogee,barbuy2018chemodynamical}.

Near-infrared spectroscopy has been decisive to extend chemical mapping across the dust-obscured inner Galaxy. APOGEE $H$-band spectra analyzed with ASPCAP provide homogeneous metallicities and abundances for large samples, revealing a metallicity distribution function MDF with multiple components whose relative weights vary with Galactic latitude \citep[e.g.][]{perez2018bulge,rojas2020many, nidever2015data}. At the metal-poor end, targeted efforts have identified genuine inner-Galaxy giants with enhanced $[\alpha/{\rm Fe}]$ and orbits confined to the central few kiloparsecs, consistent with early, intense star formation in the deepest part of the potential \citep[e.g.][]{howes2015extremely,howes2016embla,arentsen2020pigsI,duong2019herbs1,duong2019herbs2,lucey2022combs,razera2022abundance}. 
Recent analyses of APOGEE data have further shown that a population of extremely metal-poor stars exists below the metallicity limits of the ASPCAP pipeline, including objects with orbits confined to the inner Galaxy \citep{montelius2025metal}.
Cosmological models likewise predict that a substantial fraction of the oldest stars should now reside in the bulge region \citep[e.g.][]{tumlinson2009chemical,starkenburg2017pristine,el2018most, diemand2005distribution}. The metal-poor tail of the bulge/inner Galaxy has not yet been studied in much detail since the number of metal-poor stars is extremely small compared to the more metal-rich stars that dominate in the inner Galaxy. Larger samples of metal-poor inner Galaxy stars are needed to disentangle this complicated area of the
Galaxy. Additionally, the contribution of disrupted GCs to the metal-poor inner Galaxy is currently poorly constrained.

A persistent difficulty is that pencil-beam sightlines toward the bulge mix structurally distinct populations; chemical information alone cannot robustly disentangle bar-supported bulge stars from inner-disk, thick-disk, or halo interlopers \citep[e.g.][]{zoccali2008metal,ness2016apogee,perez2018bulge,lucey2022combs,razera2022abundance}. With Gaia, full 6D phase space can be combined with barred potentials to compute apocenters, vertical amplitudes, and integrals-of-motion proxies, enabling explicit chemo-dynamical classifications \citep[e.g.][]{ness2013argos,ness2016apogee,rojas2020many,lucey2022combs,razera2022abundance}. This approach isolates stars whose orbits are truly confined to the inner few kiloparsecs and thus provides a clean context to interpret abundance sequences.

In this work we exploit that strategy using the bulge Cluster APOgee Survey (CAPOS; \citet{geisler2021capos}) and the Pristine Inner Galaxy Survey (PIGS; \citet{arentsen2020pigsI}). CAPOS was primarily designed to study bulge globular cluster (GC) stars. However, given the compact size of of bulge GCs and fiber collision limitations, most fibers were available for objects outside of the GCs. To optimize  the science return, CAPOS targeted hundreds of bulge field giants per plate as well. In addition to normal, typically metal-rich bulge field stars, CAPOS also  targeted metal-poor candidates from the Pristine Inner Galaxy Survey (PIGS) surveys \citep[][]{starkenburg2017pristine, arentsen2020pigsI}.  
It is crucial to obtain high-resolution follow-up spectra for metal-poor inner Galaxy stars, providing detailed chemical abundances combined with kinematics, which can help to disentangle different stellar populations. It is also of great interest to search for the most metal-poor stars in the inner Galaxy, which are likely to be among the oldest in the Milky Way \citep[e.g.][]{tumlinson2009chemical, starkenburg2016oldest, horta2021evidence}. PIGS supplied a photometric pre-selection that enhances the incidence of metal-poor giants at low $|l|,|b|$. We assemble a homogeneous parent list of PIGS-selected field giants observed by CAPOS and adopt APOGEE/ASPCAP DR17 parameters and abundances. Orbits are integrated in barred Milky Way potentials to assign each star to bulge--bar, thin disk, thick disk, or halo classes with propagated uncertainties. On the chemical side, we focus on Fe, the best APOGEE $\alpha$ tracers (Si, Mg), Fe-peak tracers (Ni, Mn), odd-$Z$ species (Al), and light elements (C, N); Si is used as our primary $\alpha$ proxy across the full sample, and Mg trends are emphasized for first-population (1P) chemistry as identified by the N abundance, while GC-like diagnostics are assessed via the C--N, Na--O, and Mg--Al planes.

The present work builds directly on the PIGS series, but addresses a different
and complementary question. Previous PIGS studies established the efficiency
of Ca\,\textsc{H\&K} photometric pre-selection for identifying metal-poor stars toward
the inner Galaxy, showed that these stars become dynamically hotter and less
rapidly rotating toward lower metallicity, and revealed a population of
carbon-enhanced metal-poor stars in this region
\citep{arentsen2020pigsI, arentsen2020pigsII, arentsen2021pigsIII}.
Our contribution is not a new PIGS survey description nor a new measurement
of the global PIGS CEMP fraction. Instead, we use the PIGS selection as an
efficient entry point to obtain a homogeneous high-resolution, high S/N  APOGEE/CAPOS view of metal-poor field giants toward the bulge. By combining APOGEE/ASPCAP
abundances with Gaia-based orbital classification in a barred Milky Way
potential, we isolate the subset of PIGS-selected stars whose orbits are
confined to the bulge--bar region and examine whether their $\alpha$- and
Fe-peak abundance trends are consistent with rapid in situ enrichment of the
inner Galaxy or with contamination from halo, disk, or dissolved-cluster
populations.

Multiple stellar populations provide an important interpretive context for metal-poor stars in the inner Galaxy.
Globular clusters are characterized by light-element abundance variations (e.g. C–N, Na–O, Mg–Al) that distinguish first-population (1P) stars from second-population (2P) stars enriched by high-temperature H-burning products. When clusters dissolve, 2P stars (as well as 1P) can be deposited into the field, where they may mimic a chemically peculiar metal-poor population unrelated to the normal chemical evolution of the bulge. Consequently, identifying and separating stars with GC-like abundance patterns is necessary before interpreting abundance trends as signatures of in--situ enrichment. In this work we adopt a simple nitrogen-based criterion, $[\mathrm{N/Fe}]>0.7$ \citep{aoki2007carbon, geisler2025capos}, supported by the light-element planes, to flag likely 2P stars, while the remaining sample traces the chemical evolution of the field population.

Our goals are threefold. First, we want to check on the PIGS photometric metallicity, with an eye to uncovering truly metal-poor bulge field stars. 
\citet{geisler2025capos} carried out a quantitative comparison of the metallicity distribution (MD) of their sample of bulge GCs based on high SNR APOGEE spectra to that of bulge field stars, selecting the recent
large APOGEE sample from Rojas-Arriagada et al. (2020). The latter
find strong evidence of trimodality, with peaks at $[\mathrm{Fe/H}]$ = $+0.32$, $-0.17$, and $-0.66$. In contrast, the GC sample is strongly bimodal, with peaks at  $-0.45$ and a dominant peak at -1.08. The minor BGC peak falls between the two more metal-poor field star peaks, but the main difference is that the fraction of field stars below $-1$ is very
small. It is unclear why the field and GC MDs are so distinct. We hope to shed light on this conundrum, in particular by exploring the metal-poorer bulge field population. 
Second, we establish the $[\mathrm{Si/Fe}]$ and $[\mathrm{Mg/Fe}]$ sequences of a dynamically defined field-bulge sample in the metal-poor regime, quantifying slopes and dispersions over fixed metallicity windows. Third, we contrast these bulge trends with the small inner-halo tail present in our selection. The thin/thick-disk subsets are too sparse for quantitative analysis and are only acknowledged. In addition, we evaluate Fe-peak behavior (Ni, Mn) as complementary clocks, and we test for multiple-population chemistry via the light-element planes together with a simple nitrogen-based 1P/2P tagging. By combining homogeneous APOGEE abundances with explicit orbital confinement, we aim to clarify whether metal-poor field bulge stars follow the same high-$[\alpha/\mathrm{Fe}]$ sequence as nearby high-$\alpha$ populations, or retain signatures of distinct inner-Galaxy enrichment \citep[cf.][]{schiavon2017chemical,duong2019herbs1,duong2019herbs2,lucey2022combs,razera2022abundance}.

The paper is organized as follows. Section~\ref{sec:data} describes the spectroscopic dataset, PIGS selection, reductions, and working sample. Section~\ref{sec:orbits} presents the barred-potential setup and the chemo-orbital classification. Section~\ref{sec:methods} details our statistical summarization and diagnostics. Section~\ref{sec:results} reports abundance trends, light-element planes, and internal gradients within the bulge-confined locus. 
Section~\ref{sec:discussion} relates our results to the PIGS literature and interprets them in the context of inner-Galaxy formation, and Sect.~\ref{sec:conclusions} summarizes our conclusions.

\section{Data and sample definition}
\label{sec:data}

\subsection{Spectroscopic dataset and parent sample}
\label{sec:data_parent}

Our work is based on CAPOS \citep{geisler2021capos}, which obtained high-resolution near-IR spectroscopy with APOGEE for targets toward the inner Galaxy. Besides globular clusters, CAPOS deliberately sampled field giants along the bulge--bar sightlines in order to trace the chemical evolution of the central Milky Way with minimal extinction biases. We draw our parent list from APOGEE DR17, adopting the officially released radial velocities, combined visit spectra, and ASPCAP stellar parameters/abundances delivered in a homogeneous and well-documented reduction/analysis framework \citep[e.g.][]{ahumada202016th,majewski2017apache,perez2016aspcap,perez2018bulge}. 

In this paper we focus on the bulge field population selected exclusively from the Pristine Inner Galaxy Survey (PIGS) \citep{arentsen2020pigsI} and subsequently observed by CAPOS with APOGEE. Our parent list therefore consists only of PIGS-selected field-giant candidates (metal-poor prioritized) that received APOGEE fibers within the CAPOS program. This yields a field star-only sample by construction—no CAPOS cluster targets are analyzed here—concentrated at low Galactic longitudes/latitudes toward the inner few kiloparsecs of the Galactic center and spanning a broad metallicity range owing to the PIGS Ca\,\textsc{H\&K}-based pre-selection.

\subsection{Metal-poor candidate selection from PIGS}
\label{sec:data_pigs}

Our parent list comes from the PIGS survey, which uses narrow-band Ca\,\textsc{H\&K} photometry to identify metal-poor candidates through metallicity-sensitive color indices calibrated against broad-band photometry and spectroscopic reference sets \citep{arentsen2020pigsI}. CAPOS leveraged PIGS to efficiently pre-select candidate metal-poor bulge--field giants among the otherwise metal-rich inner-Galaxy population, complementing CAPOS cluster and ancillary selections. In practice, PIGS candidates were prioritized in APOGEE field designs when their colors and quality flags indicated robust low-metallicity likelihoods and giant-like loci, with additional quality controls (photometric consistency, crowding/flag checks) to mitigate contamination from dwarfs and highly reddened outliers. This strategy increases the incidence of metal-poor stars in our bulge sightlines while preserving a broad sampling in ($l,b$) across the bar region. 

We emphasize that the role of PIGS in the present analysis is target
pre-selection. The original PIGS spectroscopic metallicities and carbon
abundances were derived from low/intermediate-resolution AAT spectra and
were optimized for efficiently identifying metal-poor candidates in the inner
Galaxy \citep{arentsen2020pigsII,arentsen2021pigsIII}. In contrast, the
present work relies on APOGEE/CAPOS high resolution $H$-band spectra and the homogeneous
DR17 ASPCAP abundance scale to study detailed abundance trends and orbital
membership for the high-resolution CAPOS subset.

\subsection{Observations and reductions}
\label{sec:data_obsred}

All spectra were obtained with the APOGEE  $H$-band spectrographs ($R\sim22{,}500$) for targets lying within the innermost $\pm$10$^{\circ}$ in latitude and longitude around the Galactic center, and not planned to be observed with APOGEE-2S as part of the main SDSS-IV APOGEE survey \citep{majewski2017apache, wilson2019apache}. The CAPOS observing strategy (field layout, targeting, visit scheme, and exposure times) is described in detail in CAPOS\,I \citep{geisler2021capos}, to which we refer for a full account. In the present work we did not perform ASPCAP-independent reductions: we rely on the standard SDSS–APOGEE pipeline products. We used the frames processed by the APOGEE reduction pipeline (detector calibrations, extraction, wavelength solution, telluric and sky subtraction, per-visit RVs, and coaddition into a single high-S/N \texttt{apStar} spectrum) \citep{nidever2015data}. Stellar atmospheric parameters and chemical abundances were then derived by ASPCAP via global spectral fitting to synthetic grids, and we adopt the recommended calibrated stellar parameters and abundances provided in APOGEE DR17. \citep{perez2016aspcap,perez2018bulge, meszaros2013calibrations, jonsson2018apogee}. Throughout, we use the publicly released DR17 products \citep[e.g.][]{abdurro2025euclid, majewski2017apache, jonsson2018apogee, meszaros2013calibrations}.

\subsection{Stellar parameters, abundances, and working sample}
\label{sec:data_aspcap_clean}

We adopt the calibrated atmospheric parameters and individual abundances from APOGEE/ASPCAP (DR17), focusing on the following elements: Fe, C, N, O, Mg, Al, Si, Mn, and Ni. In the very metal-poor regime, the APOGEE $H$-band contains only a limited number of measurable Fe lines, and the ASPCAP pipeline may return unreliable or null metallicities near the edge of its grid, which reaches down to $[\mathrm{Fe/H}] = -2.5$ dex \citep{montelius2025metal,montelius2026}. In this regime $\alpha$-elements such as Mg and Si remain measurable and therefore provide more robust tracers of chemical enrichment than Fe. Throughout, Mg and Si trace the $\alpha$ sequence, while Mn and Ni represent Fe-peak behavior.

The parent list consists of the 136 PIGS-selected field giants observed by CAPOS. This is the input set for the orbital analysis in Sect.~\ref{sec:orbits}, from which a subset is subsequently assigned to the bulge--bar component by our chemo–dynamical classification. We apply a single uniform requirement on the combined APOGEE spectrum, ${S/N}\ \ge\ 50$, which ensures adequate abundance precision for metal-poor giants while retaining statistical power. After this cut, the working sample contains $N_{\rm tot}=122$ stars. The nominal ASPCAP abundance uncertainties, together with the empirically calibrated $[\mathrm{Fe/H}]$ uncertainties
described in Sect.~\ref{sec:methods_abund}, are included in the online catalog.

\subsection{Phase-space information and orbital overview}
\label{sec:data_orbits_overview}

For each star in our CAPOS–PIGS working sample we compile the phase–space inputs from uniform sources: equatorial coordinates (RA, Dec) and proper motions $(\mu_{\alpha},\mu_\delta)$ from Gaia DR3 \citep{prusti2016gaia, vallenari2023gaia}; a fixed heliocentric distance given by the PIGS median estimate \texttt{DIST50} \citep{arentsen2020pigsI}; and heliocentric radial velocity $V_{\rm HELIO}$ with its uncertainty \texttt{VERR} from APOGEE/ASPCAP (DR17). 

These inputs are used to compute stellar orbits in an inner–Galaxy Milky Way potential using the \texttt{gala} dynamics package \citep{price2017gala}. The integrations are performed in a rotating reference frame that mimics the Galactic bar, and observational uncertainties are propagated through Monte Carlo sampling of the phase–space coordinates. From the ensemble of realizations we derive characteristic orbital quantities such as apocenter, pericenter, maximum height above the plane, eccentricity, angular momentum, and the Jacobi integral. These parameters allow us to distinguish stars associated with the bulge--bar from those belonging to the thick disk, thin disk, or halo. The detailed description of the adopted Galactic potential, the orbit-integration procedure, and the criteria used to assign stars to the different Galactic components is presented in Sect.~\ref{sec:orbits}.

In the present section we simply note that our final ``bulge field'' sample is defined as those stars whose orbits are consistent with confinement to the inner Galaxy and with the spatial and kinematic properties expected for the bulge--bar component, after identifying obvious disk and halo interlopers on the basis of their orbital properties. This selection yields a final sample of 98 stars.

\subsection{Construction of the DR17 orbital reference sample}
\label{sec:dr17_reference_sample}

To derive data–driven orbital boundaries for the inner Galaxy, we first built a large reference set from the public APOGEE DR17 catalog combined with StarHorse distances.  
Stars were required to satisfy: (i) ${\rm S/N}\ge80$; (ii) clean APOGEE flags;  
(iii) finite Gaia proper motions and heliocentric radial velocities with 
$\sigma_{\rm RV}\ge0$;  
(iv) StarHorse distances with relative precision 
$(d_{84}-d_{16})/(2\,d_{50})<0.30$;  
(v) stellar parameters within $3000\le T_{\rm eff}\le8000$\,K and 
$0\le\log g\le6$; and  
(vi) sky coverage $|l|\le40^\circ$, $|b|\le20^\circ$ to ensure that orbits 
potentially compatible with the bulge--bar were retained while keeping the 
computational cost tractable.  The adopted upper limits in $T_{\rm eff}$ and $\log g$ are relatively large, as the reference sample is designed to capture the full range of stellar parameters associated with orbits potentially compatible with the bulge--bar, avoiding biases from restrictive cuts. The scientific sample analyzed in this work is defined independently using the PIGS selection and APOGEE quality criteria, and is therefore not affected by this choice. These criteria yielded a parent sample of 34\,683 stars.

This reference set was used exclusively to determine the empirical dynamical 
limits of the bulge and disk components.  
Orbital integrations with $N_{\rm MC}=500$ realizations per star provided 
robust percentiles in $(R_{\rm apo},Z_{\max},e)$ from which we defined the 
component boundaries described in Sect.~\ref{sec:orbits}.  
Applying those boundaries to the same DR17 set produced a bulge–like 
reference subset of 18\,516 stars, which is used only as a statistical 
benchmark for the metallicity distributions shown in Fig.~\ref{fig:mdf}.

The purpose of this two–step strategy is to anchor the classification to 
the actual orbital distribution of DR17 rather than to a fixed geometric 
cut in present–day Galactocentric radius.  
Because the reference sample is magnitude– and color–limited by APOGEE, 
its metallicity distribution is intrinsically weighted toward 
intermediate–metallicity populations; this is accounted for in 
Sect.~\ref{sec:orbital_mdf_rv} when comparing with the metal–poor 
PIGS subsample.

\section{Orbital analysis}\label{sec:orbits}

We computed stellar orbits in a non–axisymmetric Milky Way potential that includes a rotating bar, in order to distinguish stars whose kinematics are consistent with the inner bulge--bar from those associated with the Galactic disks and halo. Integrations were performed with \texttt{Gala} \citep{price2017gala}, adopting Galactocentric Cartesian phase–space coordinates transformed from the observer frame using \texttt{Astropy} \citep{price2022astropy}. Throughout this section we refer to the chemo–orbital projections shown in Fig.~\ref{fig:rapo_rperi} (apocenter–pericenter, colored by $[\mathrm{Si/Fe}]$) and Fig.~\ref{fig:ej_lz} (Jacobi energy $E_J$ versus angular momentum $L_z$).

Our goal is to place the chemistry of the sample into a clear dynamical context. We use abundances to identify physically distinct stellar populations: $\alpha$-enhancement traced by [Si/Fe] separates the rapid early enrichment characteristic of the inner Galaxy from lower-$\alpha$ disk contaminants, while nitrogen abundances are used to flag second-population globular-cluster–like stars that may not represent the field population. By combining these chemical diagnostics with orbital confinement, we identify stars consistent with an in--situ bulge–-bar origin and distinguish them from halo, disk, or dissolved-cluster interlopers that merely currently lie in the cross the observed sightlines. 

In practice, the apocenter–pericenter map traces the radial extent of the orbits, while the $E_J$–$L_z$ plane provides a compact representation of binding energy and angular momentum. Coloring the maps by $[\mathrm{Si/Fe}]$ links these orbital families with the classical $\alpha$-enhanced sequence observed at low metallicity.

A small number of stars are assigned to disk components by the same orbital criteria.
In particular, one object classified as thick disk (2M17142793$-$2852203; orange symbol in Fig.~\ref{fig:ej_lz}) may appear visually offset from the main bulge--bar locus because it reaches relatively high prograde angular momentum ($L_{z,50}\simeq 790~\mathrm{kpc\,km\,s^{-1}}$) while remaining fairly bound in the adopted Jacobi proxy.
Its orbit is nonetheless fully consistent with a thick-disk interpretation: it has moderate eccentricity ($e_{50}=0.49$), a modest vertical excursion ($Z_{\max,50}=1.17~\mathrm{kpc}$), and a confined radial range ($R_{\rm peri,50}=2.28~\mathrm{kpc}$, $R_{\rm apo,50}=6.65~\mathrm{kpc}$). While its metallicity ($[\mathrm{Fe/H}]\approx -1.51$) lies within the range of both thick-disk and halo populations, its orbital properties favor a thick-disk classification rather than a halo interloper. Because the disk subsamples are very small, such cases do not affect the bulge-focused chemo-dynamical conclusions.

\subsection{Potential and integration setup}\label{sec:orbits_setup}

Orbital integrations were carried out for the DR17 reference set defined in Sect.~\ref{sec:dr17_reference_sample} (34\,683 stars), which was designed to retain all stars potentially compatible with the bulge--bar while keeping the computational cost tractable.
The gravitational potential consists of a Hernquist spheroidal bulge, a Miyamoto–Nagai disk, and a logarithmic halo, to which we add a triaxial rotating bar. The bar is oriented at $\phi=20^\circ$ with respect to the Sun–Galactic–center line and rotates with a fixed pattern speed; we test $\Omega_{\rm p}=33,\,43,\,53~\mathrm{km\,s^{-1}\,kpc^{-1}}$ as in \citet{fernandez2020dynamical}, bracketing common constraints for the Milky Way bar. Orbits are integrated for $\pm1$~Gyr with a timestep of 1~Myr in the co–rotating frame. We adopt $(R_0,z_\odot)=(8.2,\,0.025)$~kpc \citep{bland2016galaxy,juric2008milky} and $(U,V,W)_\odot=(11.1,\,12.24,\,7.25)\,\mathrm{km\,s^{-1}}$ \citep{schonrich2010local}, assuming a circular speed of $V_c(R_0)=238\,\mathrm{km\,s^{-1}}$ \citep{schonrich2012galactic}.

For each star we propagate measurement errors via Monte Carlo sampling of the heliocentric radial velocity only: we draw $N_{\rm MC}$ realizations from $\mathcal{N}(V_{\rm HELIO},\sigma_{\rm RV})$ with $\sigma_{\rm RV}=\max(V_{\rm ERR},0.1~{\rm km\,s^{-1}})$, keeping position, distance, and proper motions fixed. This RV–only perturbation provides a lower limit to the full orbital uncertainties, which can be dominated by distance and proper–motion errors in the bulge. From every realization we compute
\[
R_{\rm apo},\; R_{\rm peri},\; Z_{\max},\;
e=\frac{R_{\rm apo}-R_{\rm peri}}{R_{\rm apo}+R_{\rm peri}},
\]
and the Jacobi proxy $E_J \approx E-\Omega_{\rm p}L_z$. We report 16th/50th/84th percentiles for all derived quantities.

\subsection{Monte Carlo strategy and DR17 reference sample}

To derive dynamical boundaries in a data–driven manner we first constructed a large APOGEE DR17 orbital reference set. Starting from the DR17 catalog we applied quality and astrometric cuts ensuring reliable distances, proper motions, and radial velocities, and restricted the sample to stars compatible with the inner Galaxy volume in order to keep the computational cost tractable. This yielded 34\,683 stars, for which we performed Monte Carlo (MC) sampling with 500 realizations per star.

The choice of MC = 500 for this reference set is motivated by feasibility: MC = 1000 would double the number of integrations to $\sim35$ million orbits ($>400$ CPU hours). Tests on control subsets show that increasing from 500 to 1000 changes the median orbital parameters by $<2$--$3$\% for more than 95\% of stars, well below systematic uncertainties from the Galactic potential and distances. MC = 500 therefore provides statistically stable $p16/p50/p84$ percentiles for defining population–level thresholds.

Using these percentiles we selected stars whose orbits satisfy the derived bulge limits (Sect.~\ref{sec:empirical_limits}), obtaining a DR17 bulge comparison sample of 18\,516 stars. This set is used only for reference MDFs and not for the chemical analysis.

For the scientific sample of 136 stars—the APOGEE subset of PIGS—we adopted  MC = 1000. The higher sampling is justified because:  
(i) these stars drive the chemical conclusions;  
(ii) their small number makes the computational cost negligible;  
(iii) star–by–star membership benefits from narrower posteriors.  
Increasing to MC = 1000 reduces the width of the orbital percentiles by $\sim20$–$25$\% while leaving the medians unchanged within $<1$–$2$\%, confirming that MC = 500 is adequate for threshold derivation whereas MC = 1000 is optimal for individual classification.

\subsection{Empirical derivation of bulge and disk boundaries}
\label{sec:empirical_limits}

Rather than adopting fixed literature cuts, we derived the boundaries directly from the DR17 reference set. We defined a ``bulge–like core'' as the lowest 30\% in $|L_z|$, tracing low–angular–momentum bar orbits. From this core we adopted the 90th percentile in $R_{\rm apo}$ and $Z_{\max}$ as conservative limits:

\begin{equation}
R_{\rm apo} < 6.1~{\rm kpc}, \qquad
Z_{\max} < 2.9~{\rm kpc}.
\end{equation}

These values agree with independent studies: \citet{queiroz2021milky} used $R_{\rm apo}<6$ and $Z_{\max}<3$ kpc for APOGEE+StarHorse, while several works locate the bar–disk transition at $5$–$6$ kpc. The limit is larger than the geometric bar size ($\sim3$–$4$ kpc) because $R_{\rm apo}$ measures the maximum orbital excursion, not the instantaneous radius.

For the disks we kept classical eccentricity ranges but derived vertical limits from prograde orbits ($L_z>0$):

\begin{equation}
Z_{\max}^{\rm thin}=1.05~{\rm kpc}, \qquad
Z_{\max}^{\rm thick}=2.9~{\rm kpc}.
\end{equation}

\subsection{bulge--bar membership criteria}

Each Monte Carlo realization is classified using:

\[
\begin{array}{@{}l@{\quad:\quad}l@{}}
\mathrm{bulge\mbox{--}bar} & R_{\rm apo}<6.1~{\rm kpc}\ \&\ Z_{\max}<2.9~{\rm kpc},\\
\mathrm{thin\ disk}        & Z_{\max}<1.05~{\rm kpc}\ \&\ e<0.3,\\
\mathrm{thick\ disk}       & 1.05~{\rm kpc} \le Z_{\max}<2.9~{\rm kpc}\ \&\ 0.3\le e<0.6,\\
\mathrm{halo}              & \text{otherwise.}
\end{array}
\]

We do not impose an eccentricity cut on the bulge--bar component because bulge--bar orbits span a broad range of $e$.  Bulge membership is therefore defined by confinement to the inner Galaxy ($R_{\rm apo}$ and $Z_{\max}$), whereas eccentricity is used only to separate thin- and thick-disk populations.

From the $N_{\rm MC}$ realizations we compute membership fractions and adopt the maximum–probability component as the nominal class.

Applying these criteria to the sample, we classify $N_{\rm tot}=122$ stars into $N_{\rm bulge}=98$ bulge--bar members, $N_{\rm halo}=21$ halo stars, $N_{\rm thick}=2$ thick-disk stars, and $N_{\rm thin}=1$ thin-disk star. This corresponds to fractions of 80.3\%, 17.2\%, 1.6\%, and 0.8\%, respectively.

The clear dominance of the bulge--bar component indicates that the CAPOS--PIGS selection efficiently targets stars dynamically confined to the inner Galaxy, while a modest fraction of halo interlopers is expected given the line-of-sight sampling toward the Galactic center.

\subsection{Robustness of the orbital classification}

Repeating the full orbital analysis for $\Omega_{\rm p}=33,\,43,\,53~{\rm km\,s^{-1}\,kpc^{-1}}$ yields identical component counts and star-by-star assignments within the RV-perturbed Monte Carlo scheme. The distributions in the $(R_{\rm apo},R_{\rm peri})$ and $(E_J,L_z)$ planes remain indistinguishable over this range, indicating that the classification is insensitive to reasonable choices of the bar pattern speed. This is expected because the adopted cuts are based on orbit amplitudes rather than on resonant structures that depend more strongly on $\Omega_{\rm p}$.

As an external consistency check, we compared our median orbital quantities with those reported by \citet{ardern2024pristine} for the stars in common. The comparison was performed for 124 matched objects using the median values of the pericenter, apocenter, eccentricity, angular momentum, and Jacobi integral. Overall, the two determinations show good agreement in the relative orbital ordering. We find Spearman rank correlation coefficients of
$\rho=0.62$ for $R_{\rm peri}$, $\rho=0.90$ for $R_{\rm apo}$, $\rho=0.97$ for $L_z$, and $\rho=0.96$ for $E_J$, while the eccentricity shows a more moderate, but still significant, correlation of $\rho=0.49$. The median offsets, defined as this work minus Arentsen et al. (2020), are
$\Delta R_{\rm peri}=-0.25$ kpc, $\Delta R_{\rm apo}=-0.43$ kpc,
$\Delta e=+0.08$, and $\Delta L_z=+5.34~{\rm kpc~km~s^{-1}}$.
Thus, although our integration setup yields slightly smaller orbital radii and marginally larger eccentricities on average, the relative dynamical trends are preserved. The Jacobi integral also shows a strong rank correlation, but with a large absolute zero-point offset, as expected when comparing energies computed with different Galactic potentials and conventions. We therefore use this comparison as a validation of the relative orbital classification rather than as a one-to-one calibration of the absolute energy scale.

\begin{figure} 
\centering 
\includegraphics[width=\linewidth]{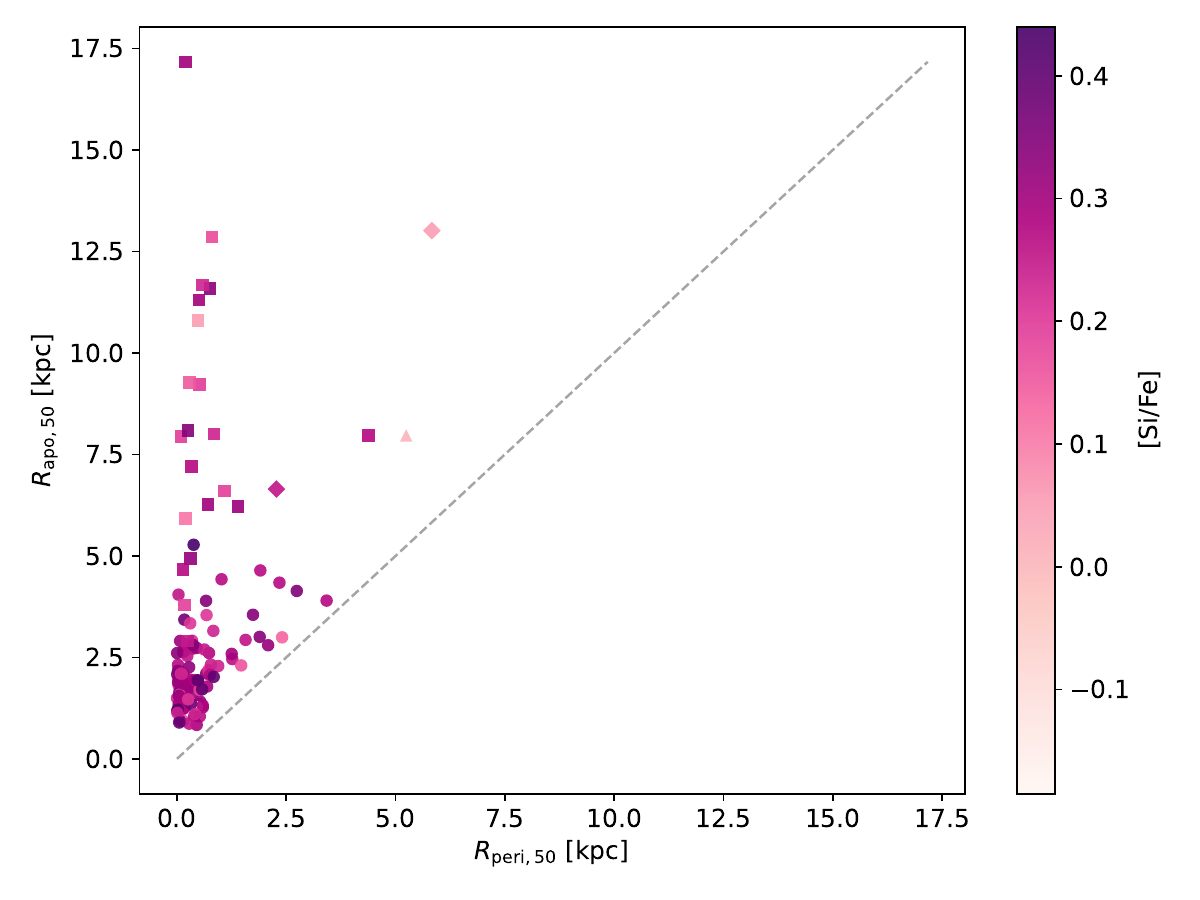} 
\caption{Apocenter ($R_{\rm apo,50}$) versus pericenter ($R_{\rm peri,50}$) for the ${\rm S/N}\ge50$ sample, color-coded by $[\mathrm{Si/Fe}]$ from APOGEE/ASPCAP. Orbital classification: bulge--bar (dot markers), halo (square markers), thick disk (diamond markers), and thin disk (triangle markers). The diagonal dashed line marks $R_{\rm apo}=R_{\rm peri}$ (circular orbits).}
\label{fig:rapo_rperi} 
\end{figure} 

\begin{figure} 
\centering 
\includegraphics[width=\linewidth]{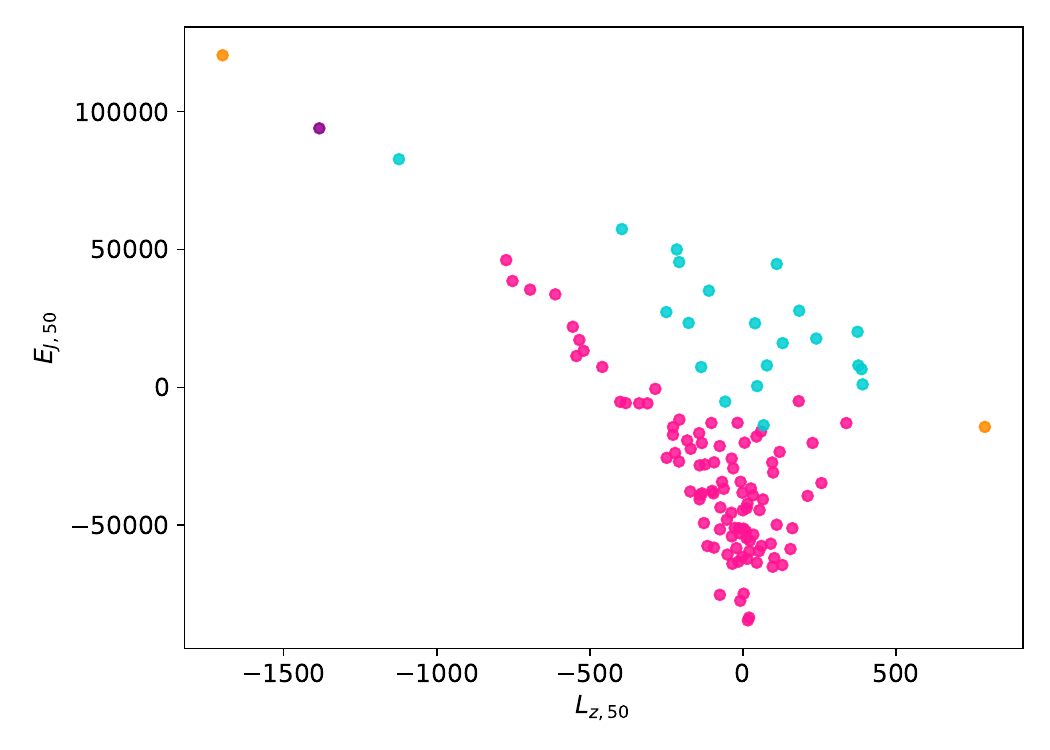} 
\caption{Jacobi-integral proxy ($E_{J,50}-\Omega_{\rm p}L_{z,50}$) versus angular momentum ($L_{z,50}$), colored by dynamical class from the chemo–orbital classification: bulge--bar (pink), halo (turquoise), thick disk (orange), and thin disk (purple).}

\label{fig:ej_lz} 
\end{figure}

\subsection{Metallicity and velocity distributions}\label{sec:orbital_mdf_rv}

\begin{figure*}
  \centering
  \includegraphics[width=\linewidth]{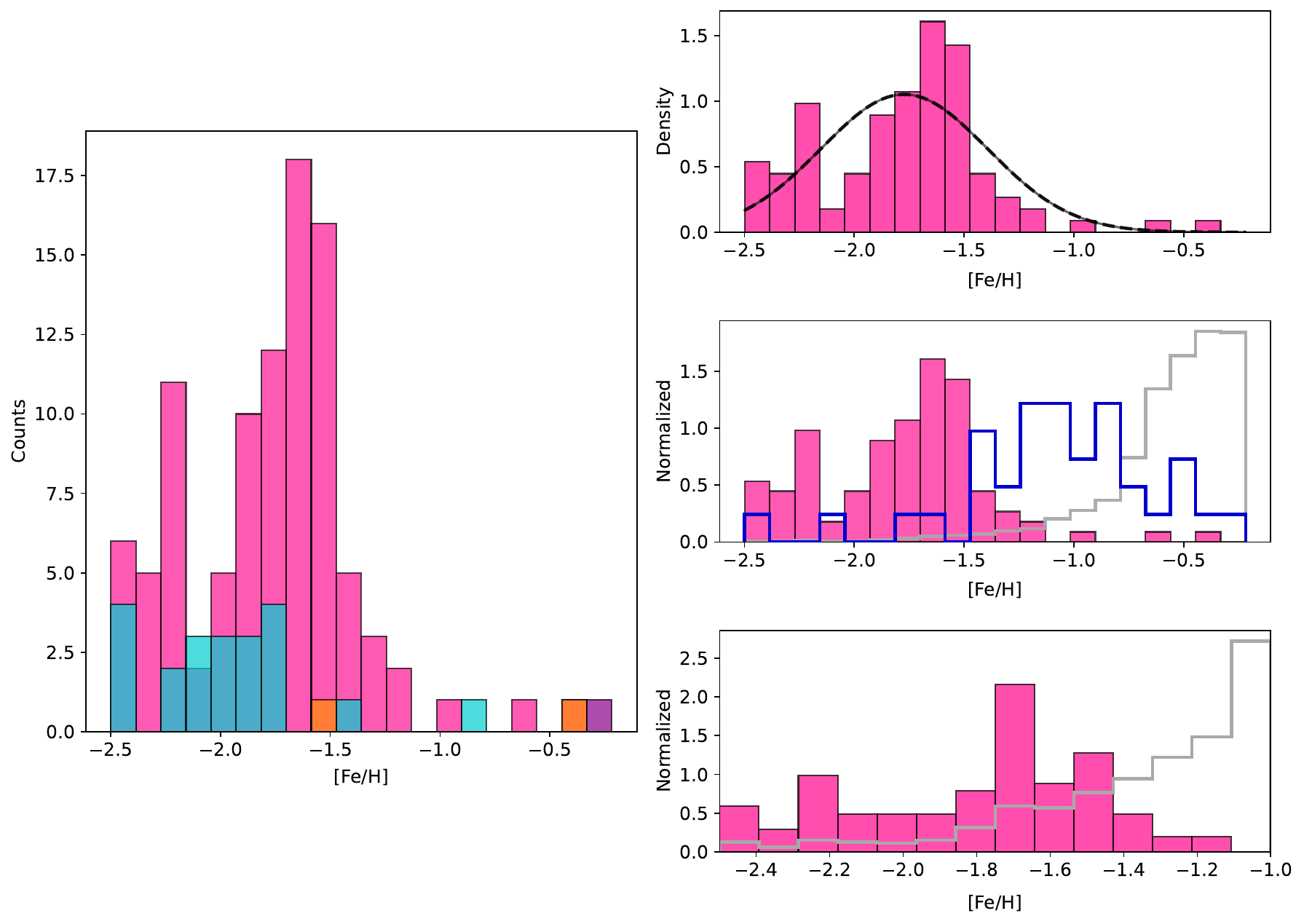}
  \caption{Metallicity distribution functions for the sample with ${\rm S/N} \ge 50$. 
Left panel: MDF in absolute counts by dynamical class, colored as: 
bulge--bar (pink, $N=98$), halo (turquoise, $N=21$), thick disk (orange, $N=2$), 
and thin disk (purple, $N=1$). 
Upper-right panel: normalized MDF for the PIGS bulge--bar subsample 
with a Gaussian Mixture Model (GMM) fit. The preferred model according to the 
Bayesian Information Criterion (BIC) is described by a component with parameters 
$G_1$: $\mu=-1.71$, $\sigma=0.38$, shown as a black dashed curve; the solid black line 
represents the total GMM reconstruction. 
Middle-right panel: comparison of the normalized MDFs for PIGS bulge sample 
(pink, $N=98$), the DR17 bulge selection (grey, $N=18\,376$), and the 
\citet{geisler2025capos} compilation of bulge globular clusters 
(blue, $N=40$).
Lower-right panel: same comparison restricted to the metal-poor regime,
including only PIGS and DR17 stars with $[\mathrm{Fe/H}] \leq -1.0$.
In this regime, the comparison includes $N=95$ PIGS stars and
$N=670$ DR17 stars.
} 
  \label{fig:mdf}
\end{figure*}

Figure~\ref{fig:mdf} displays the metallicity distribution function (MDF) by dynamical class. Qualitatively, the bulge--bar sequence is shifted to higher metallicity relative to the inner-halo tail, consistent with multi-component inner-Galaxy MDFs from microlensed dwarfs and APOGEE/ARGOS giants \citep[e.g.][and references therein]{bensby2013chemical,perez2018bulge,rojas2020many,barbuy2018chemodynamical,barbuy2025abundances}. 
The apparent truncation of both distributions near $[\mathrm{Fe/H}] \approx -2.5$ does not represent a physical limit of the population, but reflects the lower boundary of the ASPCAP spectral grid: stars below this metallicity remain observable in APOGEE spectra but cannot be reliably assigned Fe abundances by the pipeline \citep{montelius2025metal}. 

Quantitatively, a two-sample Kolmogorov--Smirnov (KS) test indicates a significant difference between the bulge--bar and halo subsamples in the present selection, with KS statistic $D=0.395$ and two-sided significance $p=0.006$. 
We summarize the distributions using the sample median and the interquartile range (IQR $\equiv Q_{75}-Q_{25}$). 
For each statistic we report the $68\%$ bootstrap confidence interval (CI), which represents the uncertainty on the estimator rather than the intrinsic spread of the stellar population.

The bulge--bar metallicity distribution has a median 
$[\mathrm{Fe/H}]=-1.71$ with a $68\%$ CI of $[-1.74,-1.68]$
and $\mathrm{IQR}=0.47$ with CI $[0.37,0.59]$ ($N=98$).
The halo subsample has median $[\mathrm{Fe/H}]=-1.96$ with CI $[-2.05,-1.87]$
and $\mathrm{IQR}=0.41$ with CI $[0.29,0.54]$ ($N=21$).
We emphasize that this value reflects the small halo tail present in our PIGS-selected sample, rather than the global Galactic halo metallicity distribution. In APOGEE-based studies, the halo MDF is typically found to peak around $[\mathrm{M/H}]\sim -1.5$, with an additional more metal-poor component near $\sim -2.1$ \citep{fernandez2016chemical}. Our halo subsample is therefore better interpreted as tracing the metal-poor tail of the halo distribution rather than the canonical halo as a whole.

Given the very small thin/thick-disk subsets (e.g. thin disk $N=1$), we refrain from interpreting pairwise tests involving those classes. The measured bulge--bar versus halo contrast matches expectations once selection is confined to bar-region orbits (small $R_{\rm apo}$ and low $Z_{\max}$; e.g. \citet{duong2019herbs1,lucey2022combs,razera2022abundance}).
The orbital boundaries used to define the DR17 bulge comparison sample were derived independently of metallicity from the full reference set of 34\,683 stars (Sect.~\ref{sec:empirical_limits}). Applying these limits yields 18\,516 DR17 stars with bulge-like orbits, which form the DR17 reference MDF shown in the comparison panels of
Fig.~\ref{fig:mdf}. Because the thresholds are based solely on dynamical quantities ($R_{\rm apo}$, $Z_{\max}$, $e$, $L_z$), they do not imprint the metallicity selection of PIGS on the DR17 comparison set.

Although the DR17 sample contains a much larger number of stars overall
($N=18,376$), only a small fraction lies in the metal-poor regime. This is
illustrated in the lower-right panel of Fig.~\ref{fig:mdf}, which isolates
the comparison to PIGS and DR17 stars with $[\mathrm{Fe/H}] \leq -1.0$. In this regime, only 670 stars in the DR17 bulge-like reference sample satisfy
$[\mathrm{Fe/H}] \leq -1.0$, compared with 95 PIGS bulge stars. At progressively lower metallicities, the DR17 counts
decrease to 197, 101, 43, and 21 stars for
$[\mathrm{Fe/H}] \leq -1.5$, $-1.7$, $-2.0$, and $-2.5$, respectively,
while the corresponding PIGS counts are 83, 51, 27, and 14. Since the MDFs
are normalized independently, their vertical amplitudes should not be
interpreted as absolute number counts. Instead, they trace the relative
metallicity distribution within each sample.

This constitutes the main selection bias relevant for Fig.~\ref{fig:mdf}:
(i) the DR17 set defines the dynamical limits using a chemically unbiased
population, (ii) the PIGS--APOGEE subset is intentionally skewed toward low
metallicity, and (iii) our observed MDF comparison is therefore not a census
of the inner Galaxy but a contrast between a metal-poor targeted sample and
a largely metal-rich reference. The orbital classification itself remains
unbiased with respect to $[\mathrm{Fe/H}]$, yet the relative heights of the
normalized histograms are governed by the different metallicity selections.

\begin{figure}
  \centering
  \includegraphics[width=\linewidth]{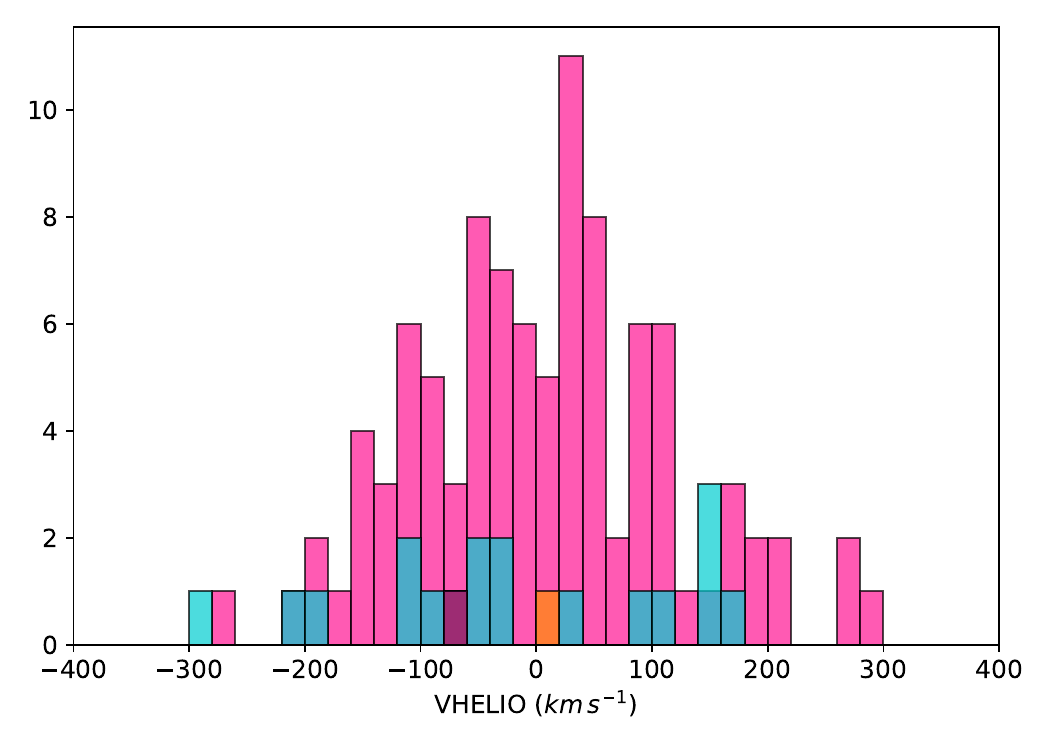}
  \caption{Heliocentric radial-velocity distributions by dynamical class for the same sample as Fig.~\ref{fig:mdf}.}
  \label{fig:rv}
\end{figure}

Figure~\ref{fig:rv} shows the heliocentric radial-velocity distributions. We find no significant bulge--bar vs halo difference at the present sample size (KS $D=0.173$, $p=0.617$). In pencil-beam lines of sight toward the bar-supported bulge, nearly cylindrical rotation and the superposition of multiple populations along the line of sight tend to dilute simple RV contrasts unless one targets specific longitudes/latitudes or works in full six-dimensional phase space \citep[e.g.][and refs.\ therein]{ness2016apogee,duong2019herbs1}. Accordingly, we regard the MDF contrast as the primary discriminator at this stage.

\section{Methods}\label{sec:methods}

\subsection{Abundance analysis and diagnostics} \label{sec:methods_abund}

We analyze element-by-element trends of $[X/\mathrm{Fe}]$ versus $[\mathrm{Fe/H}]$ for alpha tracers (Si, Mg), Fe-peak tracers (Ni, Mn), odd-$Z$ species (Na, Al), and light elements (C, N). We adopt APOGEE/ASPCAP abundances and per-element uncertainties, enforcing a single quality threshold $\mathrm{S/N}\ge 50$ (Sect.~\ref{sec:data_obsred}). 

The online table includes the nominal ASPCAP abundance uncertainties
for all stars analyzed in this work. As an additional diagnostic of the
uncertainty scale, we also provide empirically calibrated APOGEE
$[\mathrm{Fe/H}]$ uncertainties derived from an ongoing APOGEE
globular-cluster analysis (Barrera et al., in prep.). ASPCAP abundances are
derived with FERRE by fitting APOGEE spectra to synthetic spectral grids, but
the formal FERRE errors are known to underestimate the true abundance
uncertainties because systematic effects, such as imperfections in the
synthetic spectra and line-spread-function matching, are not fully captured
\citep{Holtzman_2015, GarciaPerez_2016}. APOGEE data releases therefore adopt
empirical uncertainty calibrations based on repeat observations and cluster
scatter, although these remain approximate, particularly for metal-poor stars
where the calibration data are sparse and the spectral features become weak and metallicities errors inherently larger \citep{Jonsson_2020, Mead_2024}.

In this context, Barrera et al. (in prep.) use APOGEE globular-cluster members to calibrate the random $[\mathrm{Fe/H}]$ error scale with a Bayesian model for the observed cluster metallicity distributions, assuming each globular cluster is monometallic, and thus avoiding those suspected of being metal-complex. The model uses a Student-$t$ likelihood in which the measured star-to-star
$[\mathrm{Fe/H}]$ scatter is described by the intrinsic cluster metallicity
spread compiled by \citet{Bailin_2019}, the nominal APOGEE
$[\mathrm{Fe/H}]$ uncertainty rescaled by a metallicity-dependent factor, an
additional error term depending on S/N and the formal ASPCAP uncertainty,
and residual $T_{\rm eff}$--$\log g$ trends across the cluster sequence. The
posterior calibration parameters are sampled with Markov Chain Monte Carlo
using PyMC/NUTS. The globular clusters are used only as empirical
calibrators of the APOGEE random-error scale, because their intrinsic
$[\mathrm{Fe/H}]$ dispersions are externally constrained; the inferred
correction is therefore parameterized in terms of APOGEE observables rather
than cluster-specific properties. Since the field stars lie within the
calibrated domain in $T_{\rm eff}$, $\log g$, S/N, formal ASPCAP
$[\mathrm{Fe/H}]$ uncertainty, and metallicity
($-2.5 < [\mathrm{Fe/H}] < -0.15$), this error model can be evaluated
star by star for the field sample. Overall, this increases the median $[\mathrm{Fe/H}]$ uncertainty from
$\sim0.027$ dex to $\sim0.061$ dex, corresponding to an inflation factor of
$\sim2.2$. Despite this increase, the absolute change in the uncertainty
scale is small and does not affect the interpretation of the metallicity
distribution or the abundance trends discussed in this work. These calibrated
values are listed for reference in the online catalog, while the analysis
presented here remains tied to the published ASPCAP abundance scale. This shows that even at the metal-poor tail of our sample, metallicities are well measured.

In our inner-Galaxy context, Si is treated as the most robust $\alpha$ proxy across the full sample, while Mg is emphasized mainly for first-population (1P) chemistry to mitigate known ASPCAP abundance systematics in N-rich giants \citep{perez2018bulge, geisler2025capos}. Oxygen is included in the light-element diagnostic planes but not used for formal regression fits.

For 1P/2P tagging we adopt the CAPOS convention based on nitrogen: stars with $[\mathrm{N/Fe}]>0.7$ are flagged as second-population (2P) \citep{geisler2025capos}. Also, we evaluate GC-style diagnostic planes (e.g. C–N, Mg–Al) to test for multiple-population (anti)correlations; this assessment is conducted independently of the nitrogen-based tagging, which remains the sole criterion for 1P/2P classification.

For each element $X$, we characterize $[X/\mathrm{Fe}]$ as a function of $[\mathrm{Fe/H}]$ using: (i) a running median $\widetilde{[X/\mathrm{Fe}]}$; (ii) a robust dispersion via the median absolute deviation (MAD); and (iii) non-parametric bootstrap $68\%$ confidence bands. Linear trends over fixed metallicity windows are obtained with ordinary least squares (OLS) (unweighted): we use $-2.5\le[\mathrm{Fe/H}]\le-0.4$ for Si, Mg, Na, and C, and $-2.5\le[\mathrm{Fe/H}]\le-0.6$ for Mn. Slope and intercept uncertainties are quoted as central $68\%$ bootstrap percentile intervals from $5\times10^{3}$ resamples. Distribution contrasts between dynamical classes use two-sample Kolmogorov–Smirnov tests; correlation coefficients are Spearman’s $\rho$ with two-sided $p$-values. For proportions (e.g. the 2P incidence) we report Wilson-score $68\%$ intervals. 

All abundance planes are stratified by the dynamical class derived in Sect.~\ref{sec:data_orbits_overview} (bulge--bar, thin disk, thick disk, halo). Orbital classifications are not re-computed here; we carry forward the consensus classes and, when appropriate, encode membership probabilities via symbol in the chemical plots. Because the thin- and thick-disk subsamples contain only one or two stars, they are shown for visual reference only and no quantitative chemical trends are inferred from them. Our analysis therefore focuses on the bulge--bar population and, where relevant, the halo comparison sample.

\subsection{Selection effects, biases, and implications for the MDF}

The metallicity distribution of the bulge\,bar sample is shaped by two conceptually distinct processes: 
(i) the photometric pre--selection of the Pristine Inner Galaxy Survey (PIGS), which determines which stars entered the spectroscopic follow-up, and 
(ii) the dynamical classification based on orbital parameters derived from the independent DR17 reference set.

PIGS was specifically designed to identify metal-poor stars using Ca\,H\&K narrow-band photometry \citep{arentsen2020pigsI}. 
This strategy is highly efficient for finding stars with ${\rm [Fe/H]}\lesssim-2$, but it is not metallicity-complete and intentionally over-represents the metal-poor tail of the inner Galaxy. 
Additional observational constraints, such as the magnitude ranges ($14<g<17$ or $13.5<G<16.5$) and parallax-based filters to reduce foreground disk contamination will further disfavor very distant or highly reddened giants. 
Consequently, the MDF of the PIGS--APOGEE subset should be interpreted as a targeted view of the metal-poor bulge rather than as an unbiased census of the bulge MDF.
This bias can be quantified by comparing the metal-poor fractions in the PIGS-selected sample and in the DR17 reference set. Restricting to $[\mathrm{Fe/H}] \le -1.0$, the PIGS bulge sample contains 95 stars, whereas only 670 stars in the much larger DR17 bulge-like sample (18\,516 objects) fall in this regime. This corresponds to a metal-poor fraction of $\sim 97\%$ in the PIGS sample compared to $\sim 4\%$ in DR17, illustrating the strong enhancement of the metal-poor tail introduced by the PIGS selection.

Importantly, the orbital definition of the bulge is independent of this metallicity pre--selection. 
The dynamical boundaries in $(R_{\rm apo},Z_{\max},e)$ were derived exclusively from the DR17 reference sample of 34\,683 stars (Sect.~\ref{sec:dr17_reference_sample}), which is not chemically biased by the PIGS targeting. 
Applying those limits to DR17 yields a bulge--like benchmark of 18\,516 stars used solely for statistical comparison in Fig.~\ref{fig:mdf}. 
Therefore, while our bulge\,bar MDF is chemically biased by design, the membership assignment itself is not.

A second layer of selection arises from our analysis requirements. 
The adopted $\mathrm{S/N}\ge50$ threshold removes faint spectra and could in principle bias the MDF toward higher metallicities. 
However, the median metallicity of the bulge--bar sample is insensitive to this choice:  for $\mathrm{S/N}\ge30$, $\ge40$, and $\ge50$ the same 98 stars are retained with  $\langle[\mathrm{Fe/H}]\rangle=-1.714$, while tightening to $\mathrm{S/N}\ge60$ leaves 77 stars with a virtually identical median  $\langle[\mathrm{Fe/H}]\rangle=-1.712$. 
This demonstrates that the metallicity scale is not driven by marginal-quality spectra but by the intrinsic distribution of the sample.

The orbit-based classification may also admit a small number of halo interlopers. The current dataset contains 98 bulge--bar, 21 halo, 2 thick-disk and 1 thin-disk stars, corresponding to a maximum non-bulge fraction of $\sim20\%$.  This value should be interpreted as an upper limit rather than a measured contamination level, since the probabilistic orbital assignment naturally allows partial overlap between dynamical components in the inner Galaxy.

To assess the impact of such uncertainties on the MDF shape, we performed a Monte Carlo test in which 10\% of the bulge\,bar stars were randomly replaced by metallicities drawn from the global DR17 distribution and the Gaussian Mixture analysis was repeated. 
Model selection was evaluated using the Bayesian Information Criterion (BIC), which balances goodness of fit against model complexity; lower BIC values indicate a statistically preferred description, while differences $\Delta{\rm BIC}\gtrsim6$ are generally considered significant evidence for the model with the smaller value. 
The resulting BIC values for a two-component model range between 96.7 and 107.5 across the different random realizations, remaining comparable to or larger than the single-component solution (BIC=97.1), indicating that adding an additional MDF component is not statistically justified. 
While the MDF in Fig.~\ref{fig:mdf} may appear visually bimodal, this impression is likely driven by small-number statistics and the metal-poor selection of the PIGS sample. We therefore conclude that the preference for a unimodal MDF is robust against plausible misclassification.

In summary, the MDF presented here should be interpreted as a deliberately
metal-poor view of the inner Galaxy. The selection effects mainly affect the
overall metallicity scale and the relative weights between surveys, but they
do not introduce artificial multimodality nor compromise the orbit-based
identification of bulge--bar members. The key result for the present work is
therefore not the absolute MDF of the inner Galaxy, but the existence of a
substantial sample of metal-poor stars with bulge-like orbits, extending down
to the practical ASPCAP metallicity limit near $[\mathrm{Fe/H}]\sim -2.5$.
A fuller interpretation of their origin is deferred to
Sect.~\ref{sec:discussion}.

\section{Results}\label{sec:results}

\subsection{The $\alpha$ tracers: Si and Mg}\label{sec:results_alpha}

\begin{figure}
  \centering
  \includegraphics[width=\linewidth]{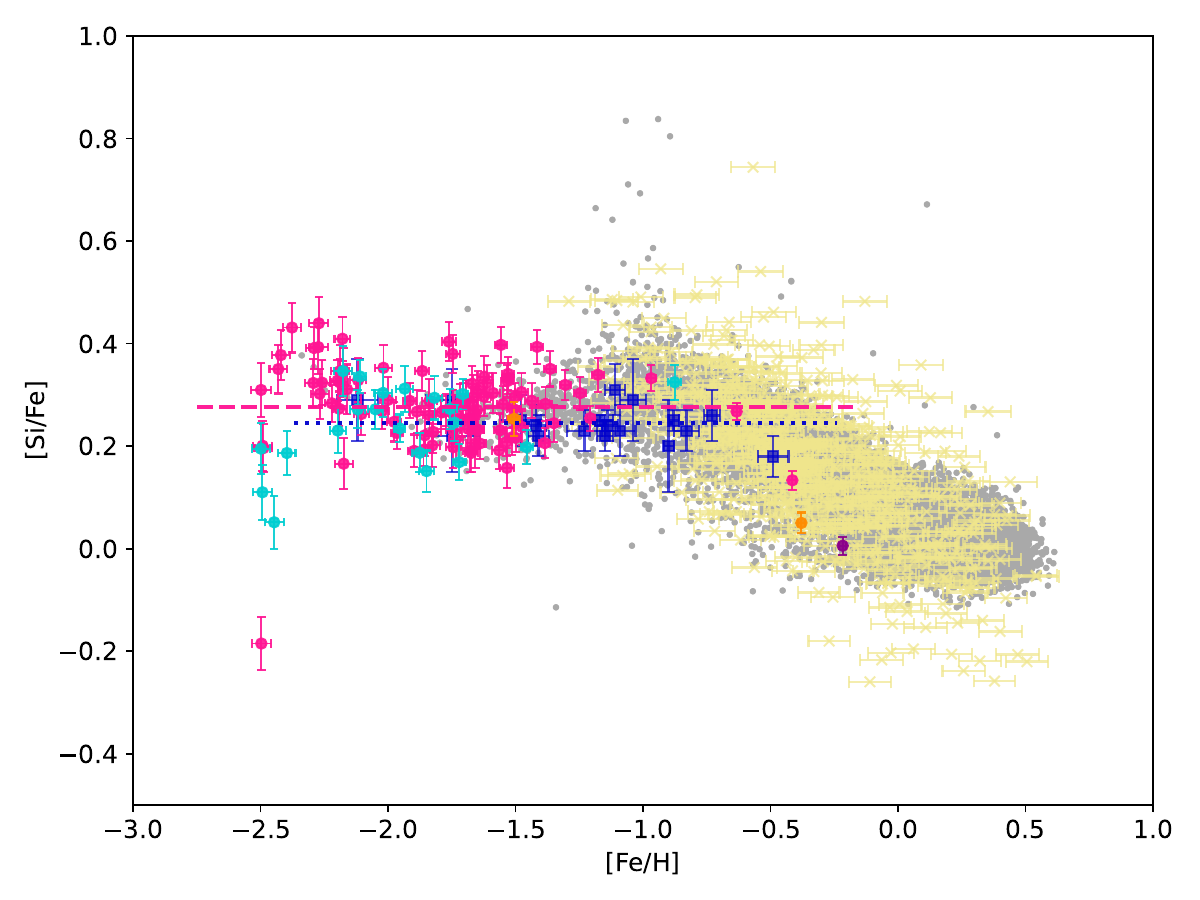}
  \caption{$[\mathrm{Si/Fe}]$ as a function of $[\mathrm{Fe/H}]$ for the sample ($\mathrm{S/N}\ge 50$), colored by dynamical class as in previous figures. Small grey points show the APOGEE DR17 distribution as a uniform background. Literature comparison samples are overplotted using fixed symbols and colors: green diamonds for \citet{bensby2013chemical}, black circles for \citet{alves2010chemical}, olive triangles for \citet{perez2013very}, khaki crosses for \citet{duong2019herbs1}, and blue squares for \citet{geisler2025capos} GC means. Horizontal lines indicate the mean $[\mathrm{Si/Fe}]$ values for the bulge--bar sample (deep pink dashed line) and the globular cluster sample of \citet{geisler2025capos} (blue dotted line). Each line is drawn over the metallicity range covered by the corresponding sample, extended by $\pm 0.25$ dex.}
  \label{fig:si}
\end{figure}

\begin{figure}
  \centering
  \includegraphics[width=\linewidth]{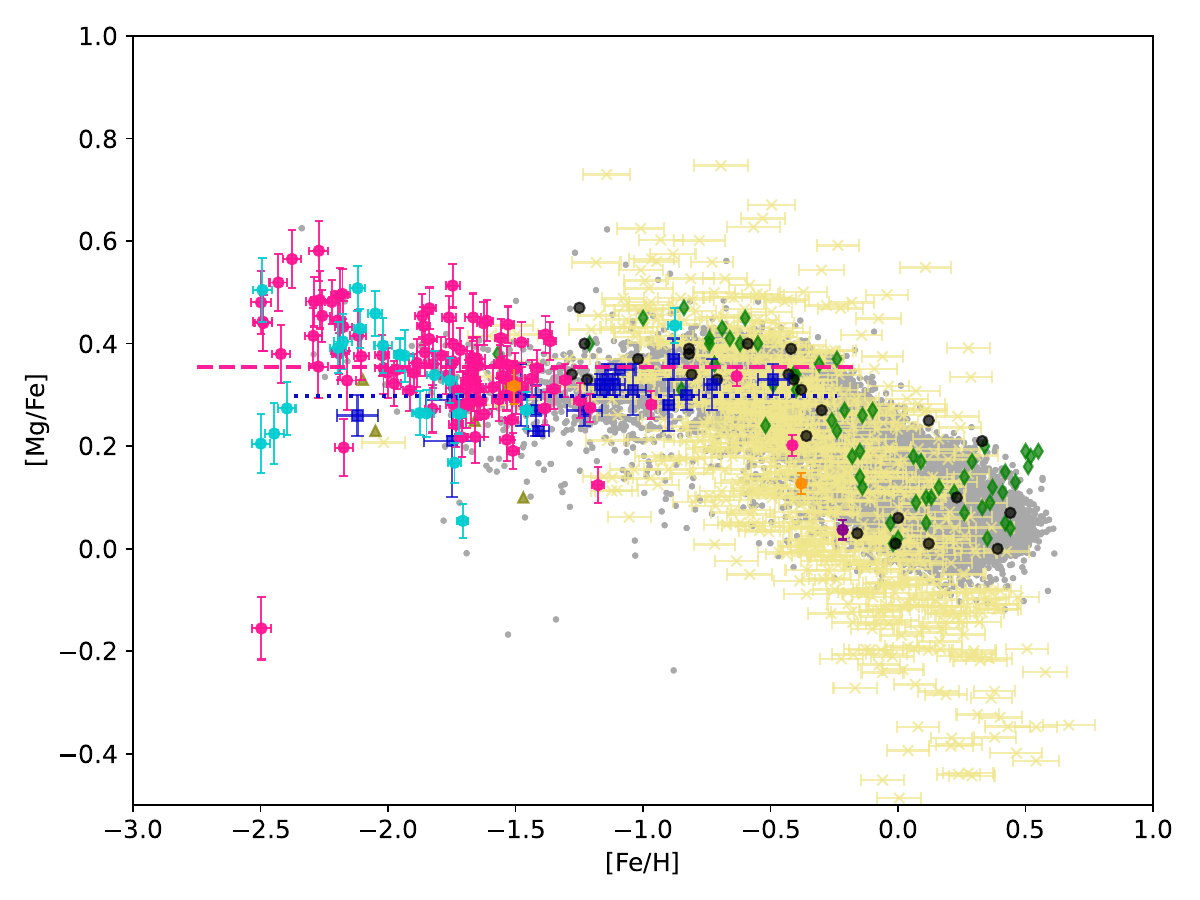}
  \caption{$[\mathrm{Mg/Fe}]$ as a function of $[\mathrm{Fe/H}]$ for the  sample ($\mathrm{S/N}\ge 50$). Colors, markers, background distribution, literature comparison samples, and mean-value lines are as in Fig.~\ref{fig:si}.}
  \label{fig:mg}
\end{figure}

Figures~\ref{fig:si} and \ref{fig:mg} display $[\mathrm{Si/Fe}]$ and $[\mathrm{Mg/Fe}]$ versus $[\mathrm{Fe/H}]$, including horizontal lines that indicate the mean abundance levels for the bulge--bar sample and for the bulge globular clusters from \citet{geisler2025capos}. 
In very metal-poor APOGEE spectra, $\alpha$-elements such as Mg and Si retain diagnostic power when Fe becomes poorly constrained \citep{montelius2026}.

Using bulge--bar stars over $-2.5\le[\mathrm{Fe/H}]\le-0.4$, we fit a straight line by ordinary least squares over the selected metallicity range. Slopes are quoted as $b^{+u}_{-\ell}$ (dex per dex), with $68\%$ bootstrap percentile CIs from $5\times10^{3}$ resamples:

\[
\begin{aligned}
\frac{\mathrm{d}[\mathrm{Si/Fe}]}{\mathrm{d}[\mathrm{Fe/H}]}
= -0.020^{+0.012}_{-0.051},
\qquad
\frac{\mathrm{d}[\mathrm{Mg/Fe}]}{\mathrm{d}[\mathrm{Fe/H}]}
= -0.097^{+0.062}_{-0.133}.
\end{aligned}
\]
Both slopes are formally negative but statistically consistent with zero, indicating an approximately flat $\alpha$ sequence across this metallicity interval.

For comparison, we repeat the same procedure for the halo subsample over the identical metallicity range:
\[
\begin{aligned}
\frac{\mathrm{d}[\mathrm{Si/Fe}]}{\mathrm{d}[\mathrm{Fe/H}]}
= +0.082^{+0.120}_{-0.039},
\qquad
\frac{\mathrm{d}[\mathrm{Mg/Fe}]}{\mathrm{d}[\mathrm{Fe/H}]}
= -0.031^{+0.032}_{-0.150}.
\end{aligned}
\]
The halo distribution shows larger dispersion and slopes broadly consistent with zero, indicating the absence of a single well-defined chemical sequence. This behavior is expected for a chemically heterogeneous population assembled from multiple enrichment histories rather than a single rapid formation episode.

The canonical high-$[\alpha/{\rm Fe}]$ plateau at low metallicity and its subsequent decline toward higher $[\mathrm{Fe/H}]$ arise naturally in chemical-evolution frameworks where early enrichment is dominated by core-collapse supernovae, with Type~Ia supernovae introducing Fe but not alpha enrichment on longer delay times and driving the downturn (“knee”) in $[\alpha/{\rm Fe}]$ \citep[e.g.][]{tinsley1979stellar,matteucci1990metallicity,mcwilliam2016chemical}. In the inner Galaxy, multiple observational probes—microlensed dwarfs, high-resolution bulge giants, and large $H$-band surveys—consistently show elevated $[\alpha/{\rm Fe}]$ plateaus and a knee located at $[\mathrm{Fe/H}]\approx -0.5$ to $-0.3$ for bulge stars \citep[e.g.][]{bensby2013chemical,ness2016apogee,barbuy2018chemodynamical,rojas2020many,barbuy2025abundances}. 
Our sample primarily probes the metal-poor plateau below this transition, explaining the absence of a resolved knee and the near-zero slopes measured over $-2.5\le[\mathrm{Fe/H}]\le-0.4$. However, only three stars in our sample have $[\mathrm{Fe/H}] > -1$. Recent results from \citet{nepal2026spheroidal} show that the metallicity distribution function of the spherical bulge, based on large samples combining Gaia RVS and APOGEE spectra, also extends to $[\mathrm{Fe/H}] < -2$, reinforcing the presence of a a small but real metal-poor tail in the inner Galaxy.

A complementary description of the plateau level is provided by the mean abundances, shown as horizontal lines in Figs.~\ref{fig:si} and \ref{fig:mg}. For the bulge--bar sample we obtain $\langle[\mathrm{Si/Fe}]\rangle = 0.277$ with a 68\% CI of $[0.269,\,0.285]$, and $\langle[\mathrm{Mg/Fe}]\rangle = 0.354$ with a 68\% CI of $[0.345,\,0.364]$. For the bulge globular cluster sample of \citet{geisler2025capos}, we find $\langle[\mathrm{Si/Fe}]\rangle = 0.244$ with a 68\% CI of $[0.237,\,0.252]$, and $\langle[\mathrm{Mg/Fe}]\rangle = 0.298$ with a 68\% CI of $[0.289,\,0.307]$. The two are broadly consistent, although this comparison should be interpreted with caution, as the field and cluster samples probe different metallicity regimes and selection functions.

One object requires special consideration. The star 2M17125370$-$2906435 shows unusually low $[\mathrm{Mg/Fe}]$ and $[\mathrm{Si/Fe}]$ at $[\mathrm{Fe/H}]\approx -2.5$ (Figs.~\ref{fig:si}--\ref{fig:mg}). Other stars at similar metallicity follow the normal $\alpha$ locus, and the orbit does not display the high energies or large excursions characteristic of accreted populations (Sect.~\ref{sec:orbits}). We note that this star lies at the extreme low-metallicity limit of the APOGEE calibration, where abundance determinations are more uncertain, and quality indicators suggest that its parameters may be less reliable. We therefore retain the star for completeness but exclude it from the interpretation of the $\alpha$ trends, treating it as an unreliable abundance measurement rather than evidence for chemically peculiar evolution.

In the APOGEE $H$ band, CN sensitivity in N-enhanced giants can affect continuum placement and propagate into correlated systematics for Mg, whereas Si relies on comparatively clean \ion{Si}{i} lines and is empirically more stable \citep[cf.][]{perez2018bulge, geisler2025capos}. Although our sample contains very few N-enhanced (2P) stars---bulge--bar fraction $0.031$ with Wilson score binomial 68\% confidence interval $[0.017,0.054]$ and halo $0.095$ with $[0.046,0.181]$ (Sect.~\ref{sec:results_light})---we adopt a conservative policy: Si is used as the backbone $\alpha$ tracer across the full sample, and Mg trends are quoted for the 1P subset ($[\mathrm{N/Fe}]\le 0.7$), following \citep{geisler2025capos} . This choice does not alter our qualitative conclusions, as the Si and Mg loci are mutually consistent within uncertainties (Figs.~\ref{fig:si}--\ref{fig:mg}).

\subsection{Fe-peak behavior: Ni and Mn}\label{sec:results_fepeak}

\begin{figure}
  \centering
  \includegraphics[width=\linewidth]{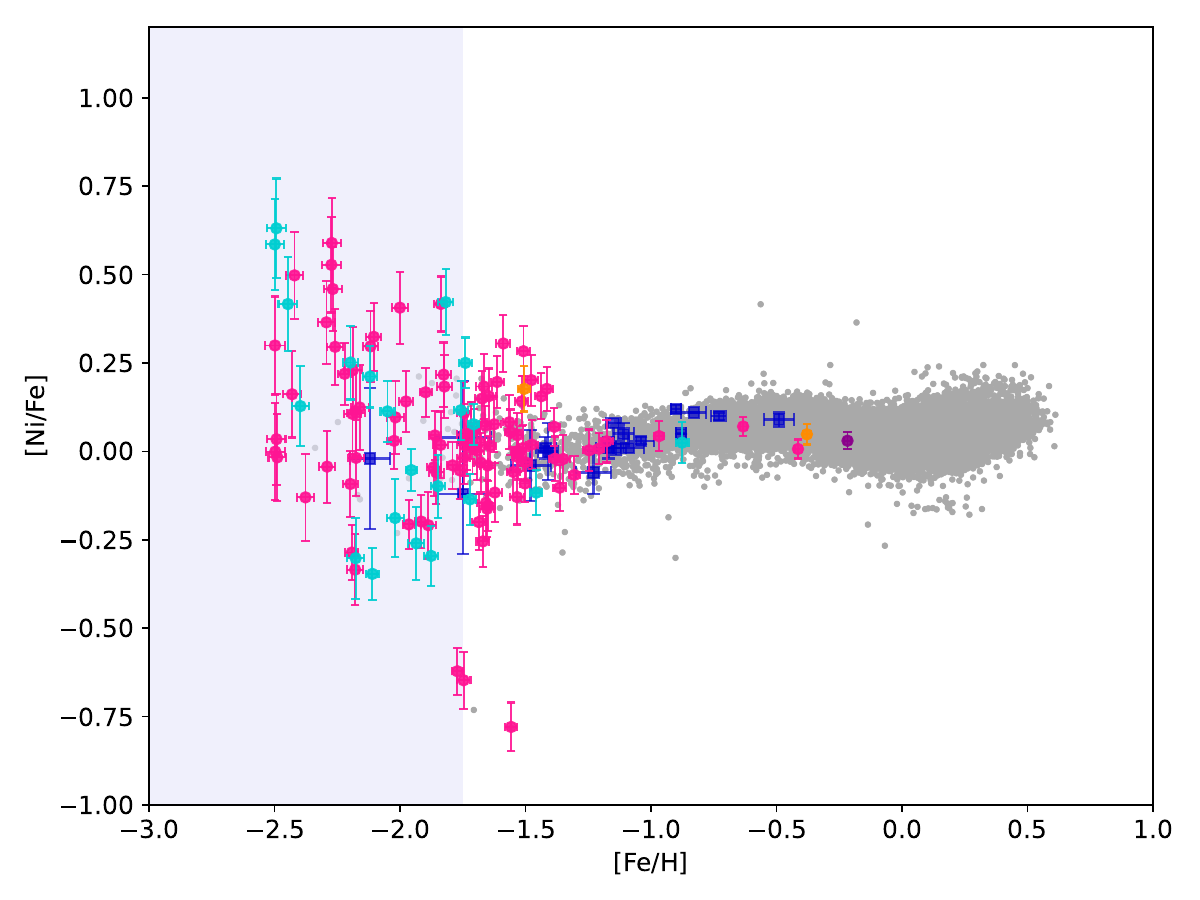}
  \caption{$[\mathrm{Ni/Fe}]$ as a function of $[\mathrm{Fe/H}]$ for the sample ($\mathrm{S/N}\ge50$), colored by dynamical class as in the previous figures. Small grey points show the APOGEE DR17 distribution as background. Blue squares correspond to the \citet{geisler2025capos} sample. Error bars are shown when available. The shaded region marks $[\mathrm{Fe/H}]<-1.75$, where [Ni/Fe] abundances should be interpreted with additional caution.}

  \label{fig:ni}
\end{figure}

\begin{figure}
  \centering
  \includegraphics[width=\linewidth]{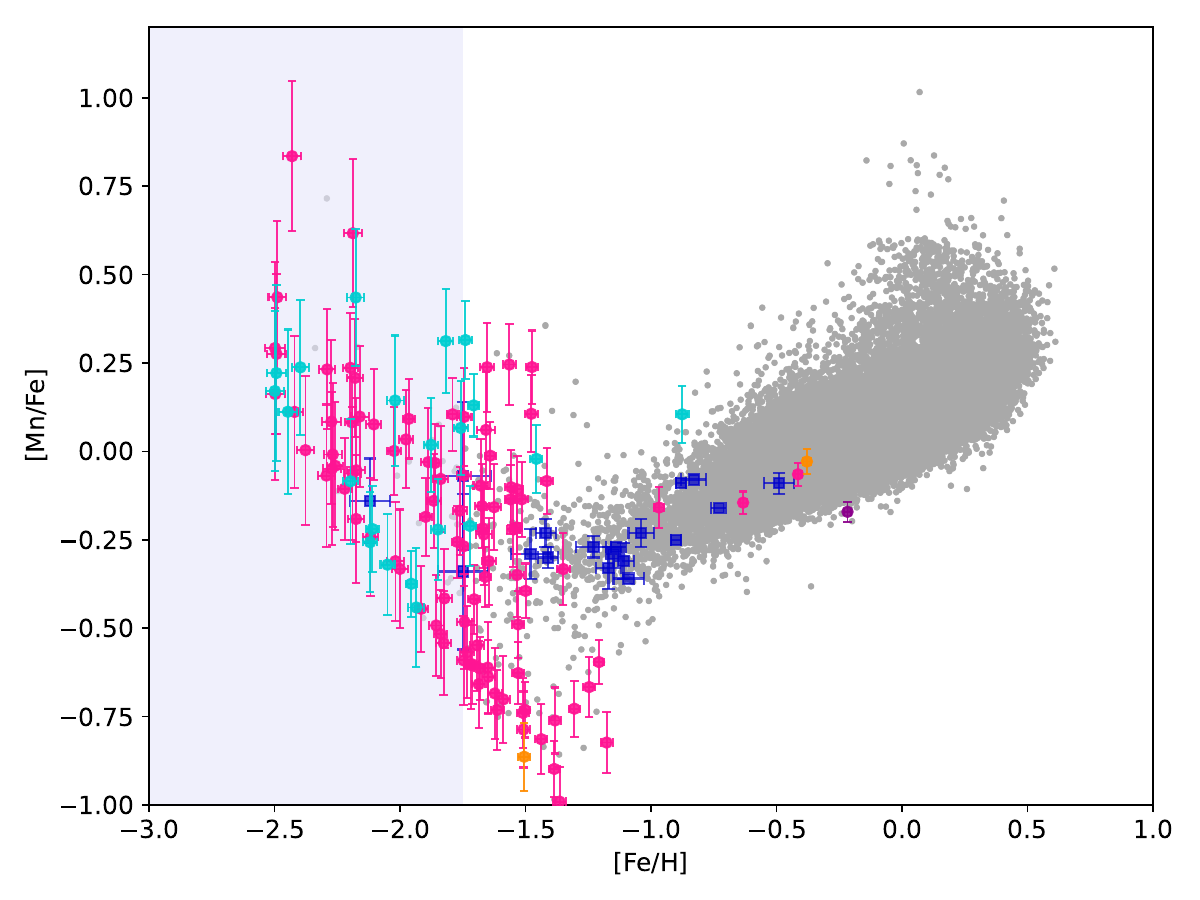}
  \caption{$[\mathrm{Mn/Fe}]$ as a function of $[\mathrm{Fe/H}]$ for the  sample ($\mathrm{S/N}\ge50$), colored by dynamical class as in the previous figures. Small grey points show APOGEE DR17 as background. Blue squares correspond to the \citet{geisler2025capos} sample. Error bars are shown when available. The shaded region marks $[\mathrm{Fe/H}]<-1.75$, where [Mn/Fe] abundances should be interpreted with additional caution.}
  \label{fig:mn}
\end{figure}

The Fe-peak tracers (Figs.~\ref{fig:ni},\ref{fig:mn}) refine the time-scale picture. Over $-1.8\lesssim[\mathrm{Fe/H}]\lesssim-1.2$, $[\mathrm{Ni/Fe}]$ remains broadly solar with modest population structure; dispersion grows toward the metal-poor end, suggesting an additional source of scatter beyond the formal abundance uncertainties. For $[\mathrm{Ni/Fe}]$ in fixed metallicity windows we obtain:
$[-1.8,-1.2]$: mean $-0.011^{+0.014}_{-0.035}$; median $+0.008^{+0.002}_{-0.0019}$.
$[-1.2,-0.8]$: mean $+0.032^{+0.037}_{+0.027}$; median $+0.029^{+0.043}_{-0.025}$.
In our data the bulge--bar stars follow the same [Ni/Fe] sequence as the halo, with no evident chemical separation between these components, and the locus agrees well with the bulge comparison samples in the overlap region.

At the lowest metallicities, the Fe-peak abundance ratios require a cautious interpretation. Although the stars satisfy our global S/N cut, APOGEE $H$-band spectra contain increasingly weak Fe-peak features toward the metal-poor end, and the ASPCAP solutions can approach the practical sensitivity limits of the grid. This effect is particularly relevant for Mn, but it can also affect Fe-peak ratios more generally. In Figs.~\ref{fig:ni} and \ref{fig:mn} we therefore highlight the region $[\mathrm{Fe/H}]<-1.75$ to remind the reader that abundance uncertainties and possible systematics become more important in this regime. The stars in this range remain central to the present work, but the detailed Fe-peak behavior there should be interpreted with additional caution rather than as a precise abundance sequence.

For Mn, the behavior is non-linear across our domain and the inferred trend depends on the limited leverage and the adopted summary statistic. An unweighted OLS fit over $-2.5 \le [{\rm Fe/H}] \le -0.6$ yields formally negative slopes for bulge--bar ($-0.586^{-0.482}_{-0.700}$) and halo ($-0.060^{+0.037}_{-0.206}$). We interpret this primarily as an empirical description of the ASPCAP [Mn/Fe] behavior in our metal-poor giant sample, noting that Mn abundances in the H band can be sensitive to analysis systematics (e.g. line strength, stellar-parameter dependencies, and S/N trends) and may not provide a clean one-parameter ``clock'' over this regime.
In particular, the apparent lower envelope in [Mn/Fe] at low metallicities is likely not physical, but instead reflects the decreasing sensitivity of Mn features in the APOGEE H-band combined with the behavior of the ASPCAP pipeline, which can effectively impose a floor in the derived abundances.
We note that our [Mn/Fe] values lie systematically below those of the comparison bulge samples at similar metallicity; this offset is plausibly related to methodological differences, including the known sensitivity of H-band Mn lines to stellar parameters and the lack of NLTE corrections in the ASPCAP scale.  
The figure suggests a shallow upturn toward the high-[Fe/H] edge of our window, qualitatively consistent with the expected rise of [Mn/Fe] at higher metallicity once SNe\,Ia contributions become important; however, our sampling above ${\rm [Fe/H]}\gtrsim -1$ is too sparse to constrain that upturn robustly within the present selection. Our interpretation of the metal-poor bulge population is therefore based primarily on the more robust $\alpha$-element tracers, especially Si, rather than on Mn.

\subsection{Odd-$Z$ behavior: Aluminum and the Mg--Al context} \label{sec:results_mgal} \label{sec:results_oddz}

\begin{figure}
  \centering
  \includegraphics[width=\linewidth]{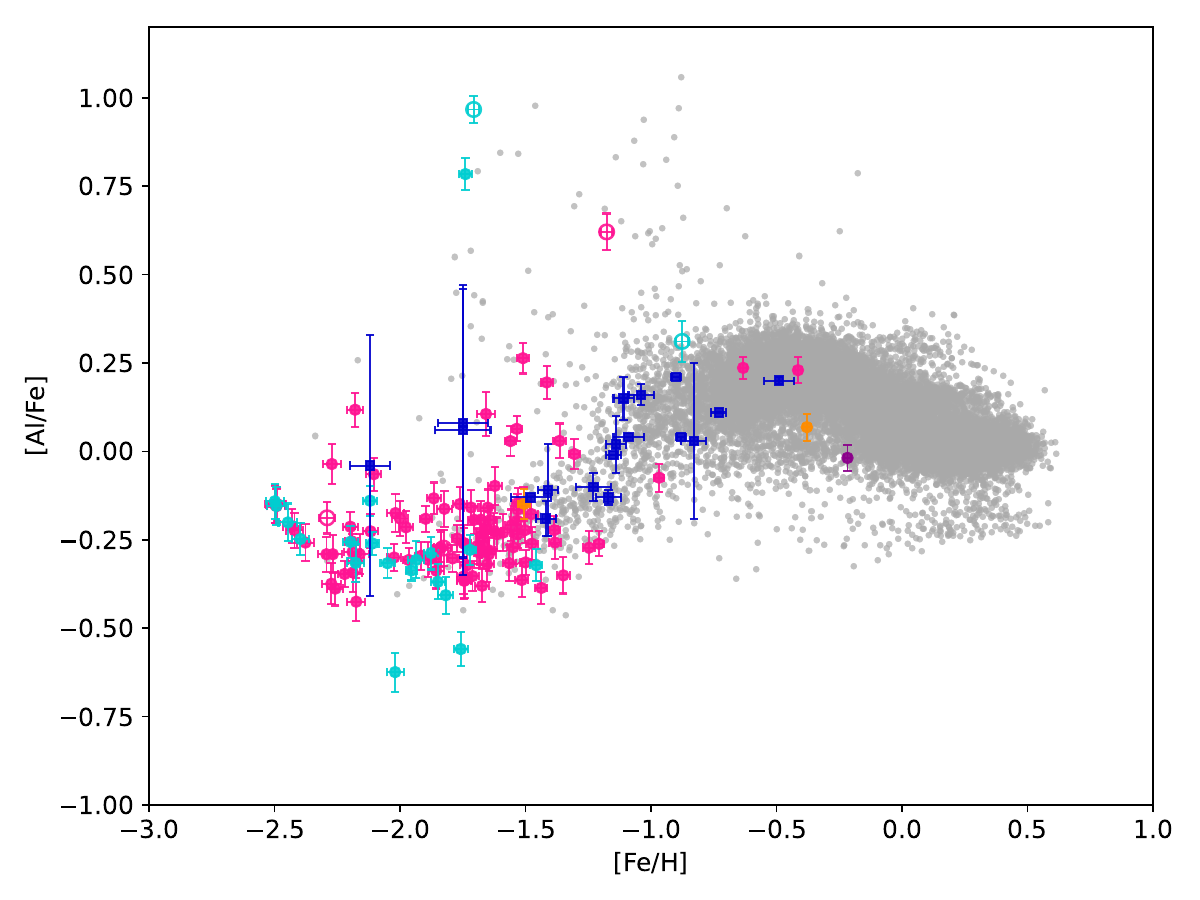}
  \caption{$[\mathrm{Al/Fe}]$ as a function of $[\mathrm{Fe/H}]$ for the same $\mathrm{S/N} \ge 50$ sample, with colors by dynamical class as in the previous figures. First-population (1P) stars are filled symbols; second-population (2P) candidates ($[\mathrm{N/Fe}]\ge 0.7$) are shown as empty dots. Small grey points indicate APOGEE DR17 background. Blue squares correspond to the \citet{geisler2025capos} sample. Error bars are shown when available.}

  \label{fig:alfe}
\end{figure}

To connect multiple-population diagnostics, we examine $[\mathrm{Al/Fe}]$ as a function of $[\mathrm{Fe/H}]$ together with the Mg--Al plane. Figure~\ref{fig:alfe} displays $[\mathrm{Al/Fe}]$ versus $[\mathrm{Fe/H}]$; symbols encode the nitrogen-based 1P/2P tagging (filled = 1P; open = 2P, defined by $[\mathrm{N/Fe}]>0.7$ - see below), while Fig.~\ref{fig:lightgrid} includes the Mg--Al panel. In globular clusters, a GC-like multiple-population signature is an extended Mg--Al anti-correlation with strong Al enhancements at fixed metallicity $[\mathrm{Fe/H}] \lesssim -1$; here we do not see such behavior. The Mg--Al locus is compact and lacks a significant anti-correlation (A Spearman rank test returns $\rho=-0.045$ with  $p=0.623$, i.e. the rank correlation is consistent with zero and there is no statistically significant monotonic association), and the bulk of the bulge--bar stars does not exhibit the large Al enhancements characteristic of GC 2P stars.

\begin{figure}
\centering
\includegraphics[width=\linewidth]{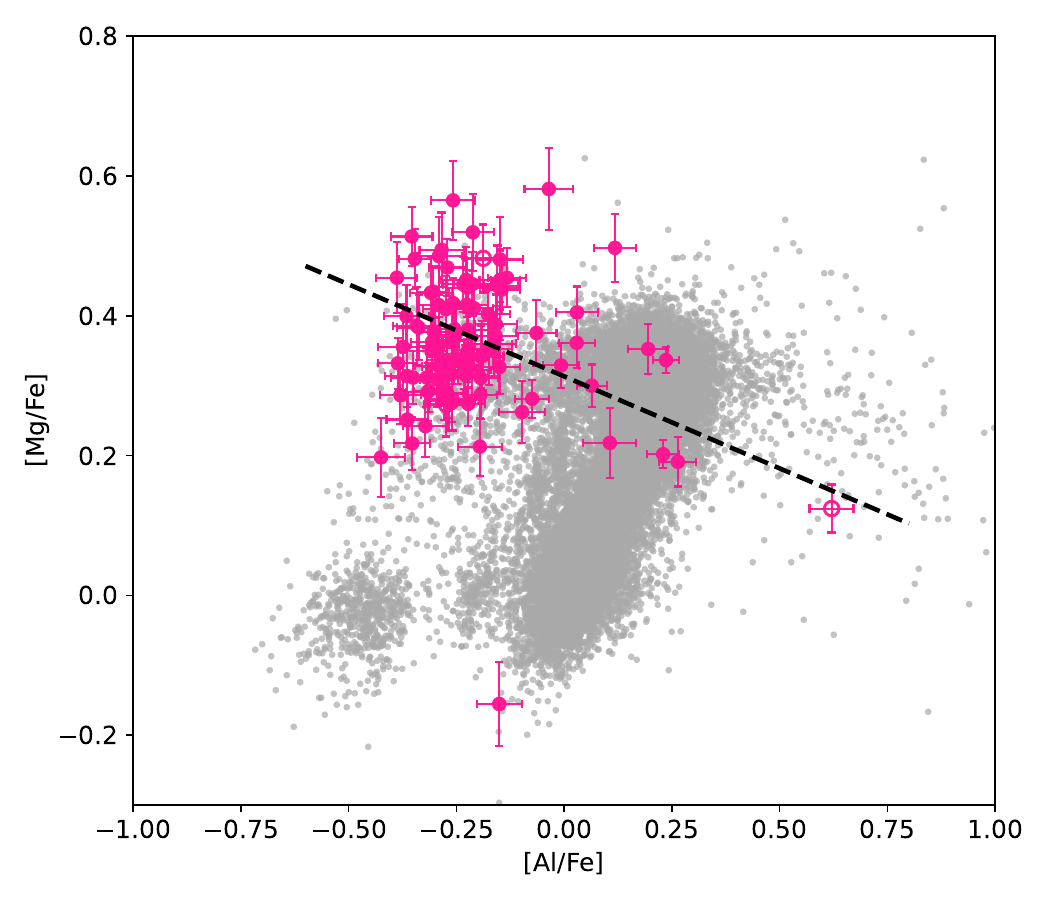}
\caption{
[Mg/Fe] as a function of [Al/Fe] for the bulge field sample (S/N $\geq$ 50). Pink symbols correspond to stars classified as bulge--bar based on the chemo--orbital analysis, while grey points represent the APOGEE DR17 parent population used as a reference background. The dashed line indicates the proposed division between in-situ and accreted stellar populations from \citet{belokurov2024situ}. Filled and empty symbols in the bulge--bar sample indicate 1P and 2P candidates, respectively.
}
\label{fig:mgal_origin}
\end{figure}

To further explore the origin of the bulge field stars, we examine the Mg--Al plane using an empirical division between in--situ and accreted populations. Figure~\ref{fig:mgal_origin} shows the [Mg/Fe]--[Al/Fe] distribution for the bulge-–bar sample (pink symbols), overlaid on the Galactic parent population from APOGEE DR17 (grey points).  We note a group of stars clustered around $[\mathrm{Mg/Fe}] \sim -0.05$ and $[\mathrm{Al/Fe}] \sim -0.5$. These objects lie in a region commonly associated with Sagittarius, although a dedicated kinematic and chemical analysis would be required to confirm this interpretation.

We include the proposed separation between in--situ and accreted stellar populations from \citet{belokurov2024situ}, shown as a dashed line. In this framework, stars located above the relation are associated with in--situ formation, while stars below the relation are typically interpreted as accreted.

We find that a significant fraction of the bulge--bar stars lie above this boundary, occupying the high-[Mg/Fe] sequence associated with rapid early chemical enrichment. This provides chemical support for an in--situ origin of the metal--poor bulge field population, in agreement with the chemo–orbital results presented in Sect.~\ref{sec:orbits}.

In our analysis, aluminum is used only as a diagnostic of possible globular–cluster (multiple–population) contamination through the Mg--Al plane. If GC-like material were present in significant fraction, stars at the metallicities we cover would presumably show large $[\mathrm{Al/Fe}]$ enhancements accompanied by Mg depletion,
producing an extended anti-correlation. Instead, the locus is compact and the majority of stars follow the field sequence, indicating that the sample is dominated by normal field enrichment.

For this reason we do not interpret $[\mathrm{Al/Fe}]$ as a tracer of the
chemical evolution of the bulge itself. Rather, we use it to identify and
control for potential multiple-population effects. Consequently, global
$\alpha$-element trends are discussed using the 1P subset (as tagged by nitrogen), which isolates nucleosynthetic evolution from any residual GC-like chemistry.

\bigskip
\subsection{Light element C}\label{sec:results_carbon}

\begin{figure}
  \centering
  \includegraphics[width=\linewidth]{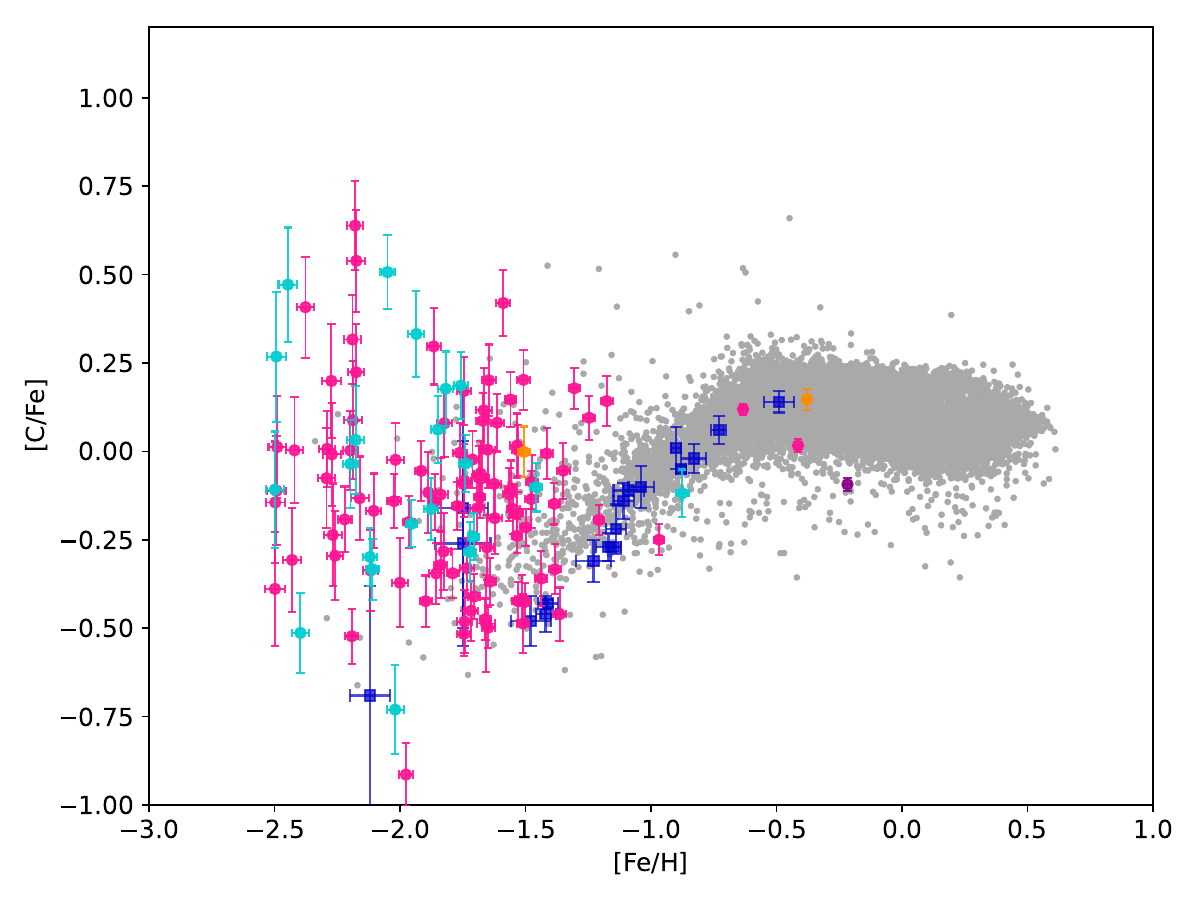}
  \caption{$[\mathrm{C/Fe}]$ as a function of $[\mathrm{Fe/H}]$ for the $\mathrm{S/N}\ge 50$ sample, colored by dynamical class as in the previous figures. Small grey points show APOGEE DR17 as background. Blue squares correspond to the \citet{geisler2025capos} sample. Error bars are shown when available}

  \label{fig:cfe}
\end{figure}

Figure~\ref{fig:cfe} shows $[\mathrm{C/Fe}]$ versus $[\mathrm{Fe/H}]$. The bulge--bar displays a shallow decline
$\mathrm{d}[\mathrm{C/Fe}]/\mathrm{d}[\mathrm{Fe/H}]=-0.038^{+0.023}_{-0.108}$,
while the halo is statistically consistent within uncertainties
$-0.087^{+0.056}_{-0.226}$.
The mild negative slope and larger scatter at low metallicity are expected in RGB samples affected by dredge-up and extra mixing; we use C as a consistency check rather than a chemo-dynamical discriminator.
Within our $\mathrm{S/N}\ge50$ sample we do not identify any carbon-enhanced star according to the  \citet{aoki2007carbon} criterion (i.e. $[\mathrm{C/Fe}]\ge +0.7$ dex, for CEMP classification). A full assessment would ideally include evolutionary (luminosity-dependent) corrections to recover natal carbon, but the present $[\mathrm{C/Fe}]$ distribution already implies a negligible CEMP incidence in this dataset.

\subsection{Light-element planes and the 1P/2P diagnostic}\label{sec:results_light}

\begin{figure}
  \centering
  \includegraphics[width=\linewidth]{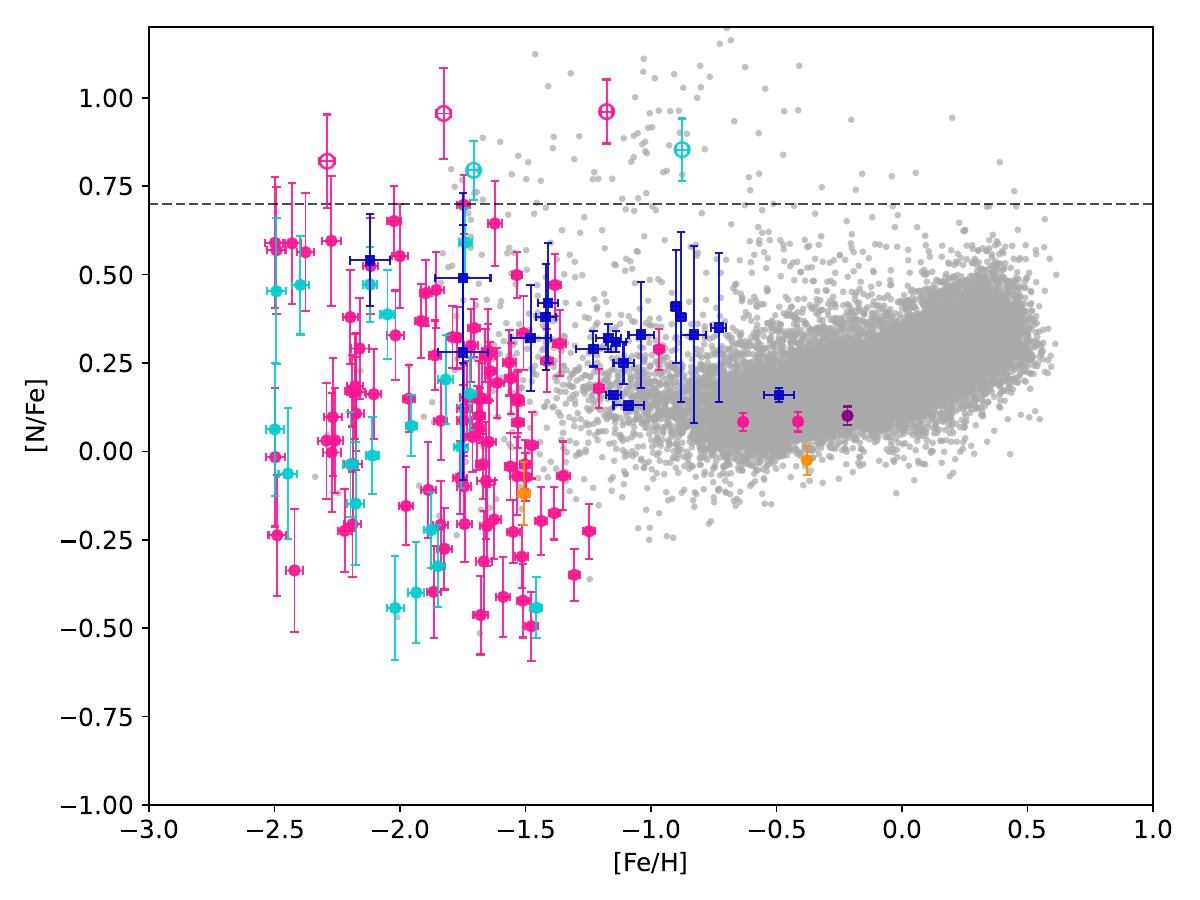}
  \caption{$[\mathrm{N/Fe}]$ as a function of $[\mathrm{Fe/H}]$ for the same $\mathrm{S/N}\ge50$ sample, colored by dynamical class as in the previous figures. The horizontal dashed line marks the 2P threshold at $[\mathrm{N/Fe}] = 0.7$; filled symbols indicate 1P, empty symbols are 2P candidates. Small grey points show APOGEE DR17 as background. Blue squares correspond to the \citet{geisler2025capos} sample. Error bars are shown when available}
  \label{fig:nfe}
\end{figure}

\begin{figure*}[t]
  \centering
  \includegraphics[width=\textwidth]{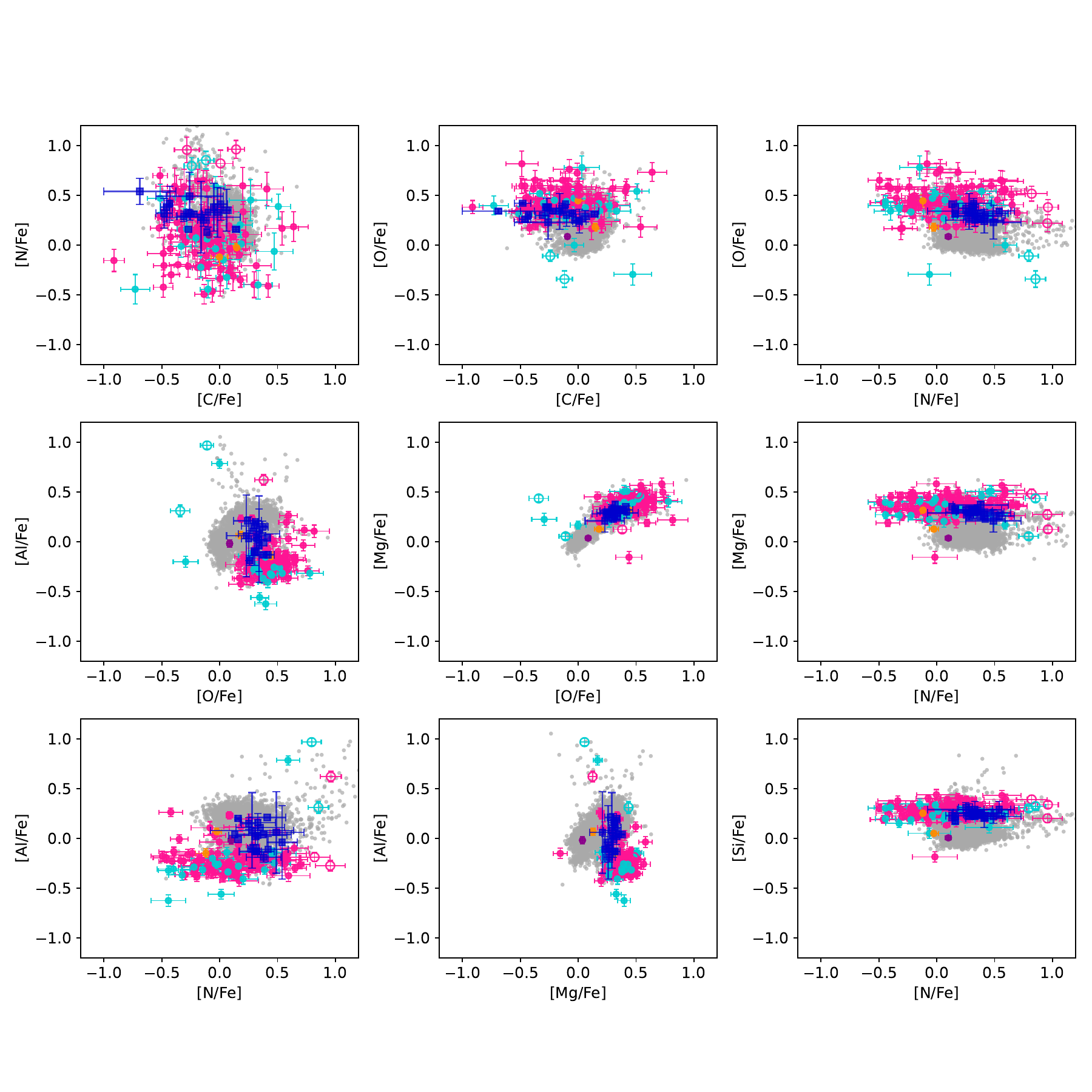}
  \caption{Light-element diagnostic planes for the $\mathrm{S/N}\ge 50$ sample, colored by dynamical class as in the previous figures. Panels (top-left to bottom-right) show $[\mathrm{N/Fe}]$–$[\mathrm{C/Fe}]$, $[\mathrm{O/Fe}]$–$[\mathrm{C/Fe}]$, $[\mathrm{O/Fe}]$–$[\mathrm{N/Fe}]$, $[\mathrm{Al/Fe}]$–$[\mathrm{O/Fe}]$, $[\mathrm{Mg/Fe}]$–$[\mathrm{O/Fe}]$, $[\mathrm{Mg/Fe}]$–$[\mathrm{N/Fe}]$, $[\mathrm{Al/Fe}]$–$[\mathrm{N/Fe}]$, $[\mathrm{Al/Fe}]$–$[\mathrm{Mg/Fe}]$, and $[\mathrm{Si/Fe}]$–$[\mathrm{N/Fe}]$. Filled symbols indicate 1P, empty symbols are 2P candidates. Small grey points mark APOGEE DR17 as background; literature overlay from \citet{geisler2025capos} with blue square markers. Error bars are shown when available). }
  \label{fig:lightgrid}
\end{figure*}

Figure~\ref{fig:nfe} shows the $[\mathrm{N/Fe}]$-based tagging, and Fig.~\ref{fig:lightgrid} summarizes the C--N, Na--O, and Mg--Al planes. The loci are compact, lacking the extended GC anti-correlations. Adopting $[\mathrm{N/Fe}]>0.7$ as the 2P threshold, we measure a low 2P incidence in the bulge--bar, $f_{2\mathrm{P}}=0.031$ with Wilson 68\% $[0.017,0.054]$ ($N=98$), i.e. only $\sim$3\% of stars show N enhancement at a level consistent with  GC-like 2P chemistry and the uncertainty on this fraction is intrinsically small. The halo tail yields a higher point estimate, $f_{2\mathrm{P}}=0.095$ with $[0.046,0.181]$ ($N=21$), but the interval is broad because of the small sample size; within these errors, the halo value is compatible with a modest 2P presence rather than a dominant dissolved-GC contribution. Consistently, rank tests in the classical planes show no significant monotonic (anti)correlations, indicating that any GC-like pattern is weak at best in this dataset.

We note that some panels of Fig.~\ref{fig:lightgrid} show modest zero-point displacements between the PIGS locus and the optical comparison bulge sample, particularly in the Mg--N, Al--N, Al--Mg, and Si--N planes. Such offsets are consistent with previously reported differences between APOGEE H-band abundances and optical analyses. Cross-survey validations have shown element-dependent systematic shifts at the $\simeq$0.1--0.15\,dex level, most pronounced for N and Al and with a measurable dependence on effective temperature \citep{jonsson2020apogee}, while residual offsets of order $\simeq$0.1\,dex are also seen for Mg and Si in comparisons with external optical references \citep{holtzman2018apogee}. These differences are understood to arise from the distinct line-formation regime in the H band, where molecular features and blending play a larger role than in the optical \citep{nidever2015data}. We therefore interpret the apparent displacements in these planes as likely reflecting methodological zero-point differences rather than intrinsic chemical inconsistencies, while the internal behavior of the PIGS sample remains self-consistent and does not exhibit GC-like anti-correlations.

Taken together, the low $f_{2\mathrm{P}}$ and the absence of strong anti-correlations reinforce a field-dominated chemical fingerprint for our inner-Galaxy sample.

\subsection{Chemo-orbital structure within the bar region}\label{sec:results_chemoorbits}

\begin{figure*}[t]
  \centering
  \includegraphics[width=\linewidth]{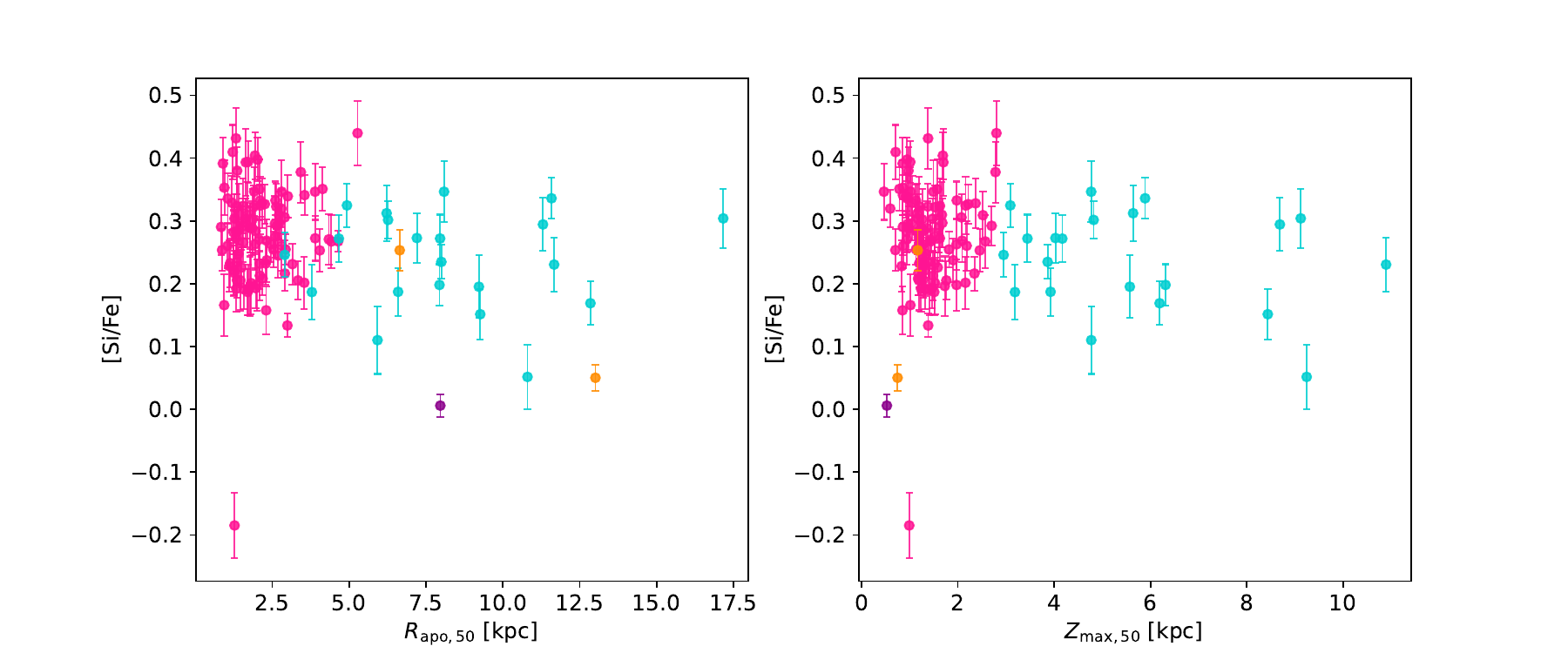}
  \caption{$[\mathrm{Si/Fe}]$ for the $\mathrm{S/N} \ge 50$ sample as a function of orbital parameters within the bar–confined locus ($R_{\rm apo}<5.7$ kpc). Left: $[\mathrm{Si/Fe}]$ vs. $R_{\rm apo,50}$. Right: $[\mathrm{Si/Fe}]$ vs. $Z_{\max,50}$. Points are colored by dynamical class as in the previous figures. Error bars are shown when available.}
  \label{fig:si_gradients}
\end{figure*}

Figures~\ref{fig:rapo_rperi}, \ref{fig:ej_lz}, and \ref{fig:si_gradients} illustrate that consensus bulge--bar members are dynamically confined (small $R_{\rm apo}$ and $Z_{\max}$; compact locus in $(E_J,L_z)$). Within this domain, and using Si as our robust $\alpha$ tracer (Fig.~\ref{fig:si_gradients}), we find no significant internal gradients in $[\mathrm{Si/Fe}]$.

To further characterize the chemical behavior of the dynamically selected bulge--bar population, we compare the mean $\alpha$-enhancement traced by Si between bulge and halo stars over the same metallicity interval ($-2.5 \leq [\mathrm{Fe/H}] \leq -0.4$). We find $\langle[\mathrm{Si/Fe}]\rangle = 0.277^{+0.008}_{-0.008}$ for the bulge--bar and $\langle[\mathrm{Si/Fe}]\rangle = 0.238^{+0.016}_{-0.017}$ for the halo, where the quoted uncertainties correspond to bootstrap 68\% confidence intervals on the mean. The halo shows a slightly lower mean value, but both populations occupy a broadly similar high-$\alpha$ locus at low metallicity, consistent with rapid early enrichment. This comparison highlights the chemical similarity between the two populations, despite their different dynamical properties.

These findings are consistent with the following gradients:

\[
\frac{\mathrm{d}[\mathrm{Si/Fe}]}{\mathrm{d}R_{\rm apo}}=+0.010^{+0.018}_{0.000}\ \mathrm{dex\, kpc^{-1}}
\]
\[
\frac{\mathrm{d}[\mathrm{Si/Fe}]}{\mathrm{d}Z_{\max}}=+0.006^{+0.022}_{-0.011}\ \mathrm{dex\, kpc^{-1}}
\]
with Spearman coefficients $\rho=\{+0.049,\,-0.112\}$ and $p=\{0.629,\,0.274\}$ for ($[\mathrm{Si/Fe}],R_{\rm apo}$) and ($[\mathrm{Si/Fe}],Z_{\max}$), respectively. The near-zero slopes and non-significant rank correlations are consistent with efficient phase mixing within the inner bulge: short orbital periods at R $\lesssim \, 3-4$  kpc and repeated pericenter passages act to erase modest chemo–spatial inhomogeneities, so that once orbits confined to the inner bulge are selected the resulting ($[\alpha/{\rm Fe}]$) field is largely uniform \citep[e.g.][]{ness2016apogee,barbuy2018chemodynamical,rojas2020many}. This dynamical homogenisation is further supported by the observed non-axisymmetric potential in the central Galaxy (bar-driven streaming and shear), which redistributes stars along inner-bulge families over many dynamical times \citep[e.g.][]{ness2016apogee,duong2019herbs1,lucey2022combs}. By contrast, the bulge–bar shows slightly higher $[\alpha/{\rm Fe}]$ than the halo, although both are consistent within uncertainties. The halo population is dynamically less confined, with larger $R_{\rm apo}$ and $Z_{\max}$, and a broader locus in the $(E_J, L_z)$ plane, consistent with stars seen in projection that are not fully phase-mixed within the inner-bulge potential \citep[e.g.][]{duong2019herbs1,lucey2022combs,razera2022abundance}.

One star (2M17125370$-$2906435) appears as a low-$[\mathrm{Si/Fe}]$ outlier in Fig.~\ref{fig:si_gradients}. 
Its chemical behavior is discussed in Sect.~\ref{sec:results_alpha}. 
From the dynamical point of view, however, the orbit is tightly confined to the inner Galaxy, with 
$R_{\rm peri}=0.46$~kpc and $R_{\rm apo}=1.28$~kpc, a moderate eccentricity ($e=0.47$), 
$Z_{\rm max}=0.99$~kpc, and low angular momentum 
$L_z=-127~{\rm km\,s^{-1}\,kpc}$. 
These properties are characteristic of bulge--bar kinematics and are incompatible with halo or Sagittarius-stream orbits. 
Therefore the star does not affect the chemo–orbital trends discussed above.

\subsection{Synthesis: origin of the \textit{field} bulge--bar population}\label{sec:results_origin}

A coherent picture emerges: (i) a significant MDF offset between bulge--bar and halo; (ii) high-$\alpha$ sequences with (slightly) declining Si and Mg across a common metallicity range; (iii) Ni broadly co-evolving with Fe and Mn showing a trend with metallicity in our ASPCAP measurements (with limited leverage at the metal-rich end); (iv) field-like light-element planes with a low 2P incidence; and (v) no internal Si gradients once bar-confinement is imposed. Taken together, these chemo–orbital signatures point to a predominantly in situ origin from the early inner disk, with secular evolution subsequently imprinting the present disk/bulge kinematics; a minor admixture of projected thick-disk/halo stars likely persists toward the most metal-poor end. A fuller interpretation is presented next.


\section{Discussion}\label{sec:discussion}

\subsection{Connection with the PIGS literature}
\label{sec:discussion_pigs}

The broader PIGS literature provides the natural context for interpreting the
CAPOS--PIGS sample, but the closest connections to our analysis are the PIGS
studies of CEMP stars, high-resolution chemical follow-up, and orbital
confinement. PIGS~III showed that CEMP stars are present in the inner Galaxy,
although the global PIGS CEMP fraction is lower than in halo samples except
at the lowest metallicities \citep{arentsen2021pigsIII}. Our analysis is not
intended to re-measure that fraction, because the CAPOS--PIGS sample is
smaller, APOGEE-targeted, and analyzed in the $H$ band. Instead, we use C, N,
and the light-element planes mainly to identify possible GC-like
second-population stars before interpreting the $\alpha$-element sequence of
the field-bulge population. PIGS~V provided high-resolution optical follow-up
for a small sample of very metal-poor PIGS stars and highlighted the need to
combine detailed chemistry and orbital information to distinguish halo-like,
GC-like, and chemically primitive inner-Galaxy populations
\citep{sestito2023pigsV}. The recent PIGS orbital analysis further showed
that many very metal-poor stars remain confined to the inner Galaxy, while
also suggesting a transition from a centrally concentrated, faster-rotating
metal-rich component to a more extended, slowly rotating metal-poor component
\citep{ardern2024pristine}. In this sense, CAPOS~XII extends the PIGS
framework from discovery and low/intermediate-resolution characterization to
a homogeneous APOGEE chemo-orbital analysis of metal-poor field-bulge stars
spanning approximately $-2.5 \leq [\mathrm{Fe/H}] \leq -0.4$.

\subsection{Implications for the origin of the field bulge}
\label{sec:discussion_origin}

The combined chemo–orbital view emerging from our analysis favors an in situ origin for the bulge tied to the inner disk, later sculpted by secular bar evolution. The Si– and Mg–abundances place the bulge--bar sequence on, or slightly above, the high-$\alpha$ thick–disk track at fixed metallicity, a configuration naturally produced by rapid early star formation followed by Type~Ia dilution. Interpreted together with our orbit-based membership, this pattern favors an inner-disk origin over a dominant classical bulge.
The stars follow bar-confined orbits with low angular momentum and
limited vertical excursions, rather than the hot, pressure-supported
kinematics expected for a merger-built spheroid. Chemically, the
continuous high-$\alpha$ sequence aligned with the thick disk and the
absence of a distinct low-metallicity knee indicate a shared
enrichment history and short formation timescale, inconsistent with
multiple independent progenitors. Together, the dynamical confinement
and chemical continuity argue that the present-day bulge field is the
secularly evolved inner disk redistributed by the Galactic bar
rather than a separate classical component
\citep[e.g.][]{melendez2008chemical,alves2010chemical,ness2016apogee,rojas2020many,babusiaux2016correlations,queiroz2021milky}.

The light-element behavior reinforces this reading. The compact C--N, Na--O, and Mg--Al planes and the low incidence of 2P-like chemistry are characteristic of field populations, not of multiple-population globular clusters. While some dissolved-cluster contribution is plausible—especially toward the most metal-poor tail—the diagnostics argue against GC debris as the primary origin of the field bulge in our sightlines \citep[e.g.][]{geisler2021capos,schiavon2017chemical,lucey2022combs}. In practice, we therefore adopt Si as the alpha tracer across the full sample and treat Mg trends conservatively on 1P stars to sidestep known ASPCAP sensitivities in N-rich giants \citep {jonsson2018apogee, geisler2025capos}. At very low metallicity, Si and Mg remain measurable in the APOGEE $H$-band while Fe lines become extremely weak, making these $\alpha$-elements more robust tracers of chemical enrichment. In particular, the Si-to-Mg ratio is sensitive to the explosion physics and nucleosynthetic yields of core-collapse supernovae \citep{montelius2026}.

The Fe-peak tracers support the same temporal picture implied by the $\alpha$ tracers. Nickel remains broadly co-evolving with iron (near-solar $[\mathrm{Ni/Fe}]$ at fixed metallicity and only modest population structure), whereas manganese shows a declining $[\mathrm{Mn/Fe}]$ with increasing $[\mathrm{Fe/H}]$ over the metallicity range sampled here. While the Mn trend is sensitive to the adopted fit window and the limited leverage at the metal-richer end of the dataset, we do not resolve strong class-to-class offsets at fixed $[\mathrm{Fe/H}]$. Overall, the Fe-peak patterns are consistent with broadly similar enrichment channels across components, with the primary differences among inner populations encoded in their $\alpha$-element time-scales rather than in distinct Fe-peak nucleosynthetic pathways.

Within the bulge-confined orbital domain, we do not resolve significant $[\mathrm{Si/Fe}]$ gradients with apocenter or vertical extent. This flat behavior is consistent with short early enrichment time-scales combined with efficient phase mixing in the inner few kiloparsecs of the bulge, which would erase modest chemo–spatial inhomogeneities while preserving the global high-$[\alpha/{\rm Fe}]$ signature. The chemically hotter and dynamically less confined tail is consistent with projected thick–disk/inner–halo interlopers, in line with orbit-cleaned inner-Galaxy samples.

Our choices were intentionally conservative: a single $\mathrm{S/N} \ge 50$ threshold, Si as the primary $\alpha$ proxy, and a consensus dynamical membership obtained by intersecting a geometric apocenter cut with a KDE selection in the Jacobi proxy. Potential systematics—abundance fitting details in cool giants, distance and proper-motion uncertainties (not propagated in the present Monte Carlo analysis), and modest-number statistics at the lowest metallicities—were monitored via bootstrap uncertainties and pattern–speed tests; the qualitative picture remains stable under these checks. 
We note that APOGEE spectra may include stars with metallicities below the ASPCAP grid limit that are returned without reliable Fe abundances. Identifying and analysing such objects requires dedicated spectral modeling beyond the scope of this work, which is restricted to stars with well-determined atmospheric parameters; their inclusion would extend the metallicity tail but would not alter the chemo–dynamical trends discussed here.
In this framework, the field bulge appears as an inner–disk population rapidly enriched at early times and subsequently rearranged by secular bar evolution, with a minority contribution from projected thick–disk/halo stars toward the most metal-poor end. Future extensions to O, Ca, Ti and selected neutron-capture tracers, combined with action–space membership and full posterior orbit modeling in barred potentials, should further refine the enrichment and mixing chronology in the central Galaxy.

\section{Conclusions}\label{sec:conclusions}
Our results can be summarized according to the following scheme:

\begin{enumerate}
\item We find a substantial population of metal-poor ($-2 \lesssim [\mathrm{Fe/H}] \lesssim -1$) stars in the bulge, extending down to $[\mathrm{Fe/H}] \approx -2.5$, the limit of the ASPCAP grid.

\item Chemo-orbital coherence: A dynamically defined bulge field sample (bar-constrained orbits) exhibits high-$[\alpha/\mathrm{Fe}]$ sequences consistent with rapid early enrichment. Within these orbits we do not resolve significant $[\mathrm{Si/Fe}]$ gradients with apocenter or height.

\item Field-like light elements: Compact C--N, Na--O, and Mg--Al planes together with a low 2P incidence argue against a dominant dissolved-GC origin for the bulk of the field bulge  especially at the metal-poor end.

\item Fe-peak timelines: $[\mathrm{Ni/Fe}]$ remains approximately solar (co-evolving with Fe) across our leverage, while $[\mathrm{Mn/Fe}]$ in our ASPCAP measurements yields an overall negative OLS slope over the analyzed range; at fixed metallicity we find no strong class-to-class offsets.

\item Formation pathway: The combined evidence favors an in situ inner-disk origin for the bulge, subsequently shaped by secular bar evolution, with a minority contribution from projected thick-disk/halo stars toward the most metal-poor tail.
\end{enumerate}

\noindent Outlook. Extending the element set (O, Ca, Ti and selected $n$-capture tracers), exploring action-space membership, and jointly modelling distance/proper-motion posteriors in barred potentials will further sharpen constraints on inner-Galaxy enrichment time-scales and mixing.
\\

\begin{acknowledgements}
We thank the APOGEE/SDSS teams for making these data possible. 
We are grateful to the referee for a constructive report that improved this paper. 
C.S. acknowledges support from ANID FONDECYT POSTDOCTORADO N°3240157.
D.G. and S.V. gratefully acknowledge the support provided by Fondecyt regular no. 1220264.
D.G. also acknowledges financial support from the Vicerrectoría de Investigación y Postgrado de la Universidad de La Serena. We would also like to acknowledge Roger Cohen for help in selecting the PIGS targets for our CAPOS fields.
This work made use of the Python scientific ecosystem, including \texttt{NumPy}, \texttt{Astropy}, \texttt{Matplotlib}, and the \texttt{gala} package for orbit integration.

\end{acknowledgements}

\noindent Based on observations obtained through the Chilean National Telescope Allocation Committee through programs CN2017B-37, CN2018A-20,
CN2018B-46, CN2019A-98 and CN2019B-31.

\section*{Data availability}

The online catalog associated with this paper contains the full list of CAPOS--PIGS stars analyzed here. It includes the APOGEE/ASPCAP DR17 quantities used in this work, namely stellar identifiers, coordinates, S/N, radial velocities, stellar parameters, abundances, and their nominal ASPCAP uncertainties. The catalog also includes the orbital parameters and adopted dynamical classification derived in this work. As an additional uncertainty diagnostic, we provide the calibrated $[\mathrm{Fe/H}]$ uncertainties estimated from a cluster-based calibration currently in preparation. The APOGEE spectra and ASPCAP products are publicly available as part of SDSS/APOGEE DR17. The online catalog will be made available at the CDS.

\bibliographystyle{aa}      
\bibliography{paper}

\end{document}